\documentclass{emulateapj}

\newcommand{\rashort}[2]{$#1^{\mathrm{h}}#2^{\mathrm{m}}$}
\newcommand{\arcsd}[2]{$#1^{\prime\prime}\!\!.#2$}           
\newcommand{\arcmn}[2]{$#1^{\prime}\!\!.#2$}             

\newcommand{\etal}{{et al.} }
\newcommand{\eg}{{\em eg: }}
\newcommand{\ie}{{\em ie: }}
\newcommand{\vs}{{\em vs }}
\newcommand{\msun}{\hbox{M$_{\odot}$}}

\shorttitle{Tunable-filter imaging of quasar fields at $z \sim 1$.}
\shortauthors{Barr et al.}

\begin{document}

\title{Tunable-filter imaging of quasar fields at $z \sim 1$. II. 
The star-forming galaxy environments of radio-loud quasars}

\author{J. M. Barr\altaffilmark{1}}
\affil{University of Bristol \\ Department of Physics, H.H. Wills Laboratory, Tyndall Avenue \\ Bristol, BS8 1TL \\ U.K.}
\altaffiltext{1}{Present address: Astrophysics Department, Keble Road, Oxford, OX1 3RH, U.K.}
\email{jmb@astro.ox.ac.uk}

\author{J. C. Baker}
\affil{University of Oxford \\ Astrophysics Department, Keble Road \\ Oxford, OX1 3RH \\ U.K.}
\email{jcb@astro.ox.ac.uk}

\author{M. N. Bremer}
\affil{University of Bristol \\ Department of Physics, H.H. Wills Laboratory, Tyndall Avenue \\ Bristol, BS8 1TL \\ U.K.}
\email{m.bremer@bristol.ac.uk}

\author{R. W. Hunstead}
\affil{University of Sydney \\ School of Physics, NSW 2006 \\ Australia}
\email{rwh@physics.usyd.edu.au}       

\and

\author{J. Bland-Hawthorn}
\affil{Anglo-Australian Observatory \\ PO Box 296, Epping, NSW 2121 \\ Australia}
\email{jbh@aaoepp.aao.gov.au}

\begin{abstract}

We have scanned the fields of six radio-loud quasars using the Taurus
Tunable Filter to detect redshifted [O\,{\sc ii}]\,$\lambda 3727$
line-emitting galaxies at redshifts $0.8<z<1.3$. Forty-seven new
emission-line galaxy (ELG) candidates are found. This number
corresponds to an average space density about $100$ times that
found locally and, at $L$([O\,{\sc ii}]) $<10^{42}$ erg s$^{-1}$
cm$^{-2}$, is $ 2 - 5$ times greater than the field ELG density at
similar redshifts, implying that radio-loud quasars inhabit sites of
above-average star formation activity. The implied star-formation
rates are consistent with surveys of field galaxies at $z\sim 1$.
However, the variation in candidate density between fields is large
and indicative of a range of environments, from the field to rich
clusters. The ELG candidates also cluster --- both spatially and in
terms of velocity --- about the radio sources. In fields known to
contain rich galaxy clusters, the ELGs lie at the edges and outside
the concentrated cores of red, evolved galaxies, consistent with the
morphology-density relation seen in low-redshift clusters.  This work,
combined with other studies, suggests that the ELG environments of
powerful AGN look very much the same from moderate to high redshifts,
i.e. $0.8 < z < 4$.

\end{abstract}


\maketitle

\label{firstpage}

\section{Introduction}

Tracing star formation over the history of the universe is crucial
for understanding the creation and evolution of galaxies. 
In global terms, the integrated star formation rate (SFR) per unit volume 
was at its peak at $z \sim 1$  (\eg Ellis \etal
1996; Lilly \etal 1996; Madau \etal 1996; Madau, Pozzetti \& Dickinson
1998 but see Cowie, Songaila \& Barger
1999)\nocite{ellis96,lilly96,madau96,madau98,cowie99}. However, 
the physical explanation for such strong redshift dependence is
still under question. In hierarchical models,  
$z\sim 1$ should also be an epoch of major galaxy-cluster assembly, with
numerous mergers and subcluster development (\eg Lacey \& Cole
1993\nocite{lacey93}; Khochfar \& Burkert 2001\nocite{khochfar01};
Murali \etal 2002\nocite{murali02}). 
The fraction of blue galaxies in clusters at $z = 1$ is 
higher than in similar systems at $z=0$ \citep{butcher84,dressler92} 
and some studies suggest that star formation is suppressed in 
the centres of galaxy clusters relative to
the field, at least out to $z \sim 0.5$ \citep{balogh97,balogh98}. 
Furthermore, the cores of galaxy clusters may show a deficit 
of active galactic nuclei (AGN) as well as strongly 
star-forming galaxies (Barr \etal 2003\nocite{barr03};
Miller \& Owen 2003\nocite{miller03}). The space density of AGN
also peaks at high redshift and declines after $z\sim 1$ 
(\eg Boyle \& Terlevich 1998). So, environment 
may strongly influence both star formation and AGN activity. 

Magnitude-limited, spectroscopic field and cluster samples 
of strongly star-forming emission-line galaxies (ELGs) become 
incomplete by $z \sim 1$. At this redshift and beyond, 
narrow-band imaging techniques are more efficient at detecting ELGs. 
Narrow-band searches have a major advantage over
magnitude-limited surveys in that they select objects 
directly on the basis of their star formation activity. 
At $z \sim 1$, ELGs typically have very faint continuum magnitudes, 
$I \gtrsim 21$ \citep{cowie97,cardiel03}, and are unremarkable 
in broad-band images -- if seen at all. Indeed,
studies have shown that magnitude-limited surveys can miss a
substantial amount of star formation even at $z = 0.4$ (Jones \&
Bland-Hawthorn 2001; hereafter JBH01)\nocite{jones01}. 
To cover the large volumes needed for surveys of star-forming galaxies
at moderate to high redshifts, tunable-filter instruments
offer an advantage over traditional monolithic filters
in being able to target a wide range of wavelengths in 
contiguous passbands.

This paper presents tunable-filter observations designed to detect
ELGs in the fields of six radio-loud quasars at $z\sim 1$ to
investigate the links between AGN activity, environment and star
formation at redshifts where AGN, star and structure formation all
peak.  Powerful AGN --- particularly radio-loud AGN --- are known to
inhabit regions of above-average galaxy density (Yee \& Green 1984;
Ellingson \etal 1991; Wold \etal 2000; Pentericci \etal 2000; McLure
\& Dunlop 2001; Venemans \etal 2002; Barr \etal 2003)
\nocite{yee84,ellingson91b,wold00,mclure01,barr03,
pentericci00,venemans02,barr03}, and may even trace the first
overdense regions to collapse. ELGs in some AGN fields have been
detected with narrow-band imaging using monolithic filters (Pentericci
\etal 2000; Hall \etal 2001; Kurk \etal 2001; Venemans \etal
2002)\nocite{pentericci00,hall01,kurk01,venemans02}, but to date
systematic searches of homogeneous samples, as is possible with
tunable filters, have not been carried out.

In this work, we target the redshifted [O\,{\sc ii}]\,$\lambda 3727$ 
emission line. The [O\,{\sc ii}] line has the benefit of being in the 
optical regime out to $z \sim 1.5$ and has been used widely as
an empirically calibrated measure of star formation rate (SFR)
(\eg Hutchings, Crampton \& Persram 1993; Hammer \etal 1997; 
Kennicutt 1998; Hogg \etal 1998; Gallego
\etal 2002; Hicks \etal
2002)\nocite{hutchings93,hammer97,hogg98,kennicutt98,gallego02,hicks02}.
Although a more direct SFR indicator, H$\alpha$ is only visible 
in the infrared at $z>0.5$. Also, [O\,{\sc ii}] is not as susceptible to 
dust or resonant scattering effects as, for example, the Ly$\alpha$ line. 

The Taurus Tunable Filter (TTF) etalon on the Taurus-2 instrument at
the Anglo-Australian Telescope (AAT) provides a Fabry-Perot-based
imaging system which allows high-efficiency, narrow-band imaging over
a range in wavelength. Spectral resolutions, $\approx 100-1000$ are
attainable from 3700\AA \ to 10000\AA \ (Bland-Hawthorn \& Jones
1998{\em a}, 1998{\em
b})\nocite{bland-hawthorn98a,bland-hawthorn98b}. With its high
throughput and narrow bands, TTF can achieve sensitivities of $\sim
10^{-17}$ erg s$^{-1}$ cm$^{-2}$ arcsec$^{-2}$ ($3\sigma$) in 12
minutes of integration. This is sufficient to probe [O\,{\sc ii}]
emission in galaxies with star formation rates greater than a few
\msun \ yr$^{-1}$ at $z \sim 1$.

Baker \etal (2001\nocite{baker01}; hereafter Paper 1) demonstrated
that it is possible to target [O\,{\sc ii}] emission from star-forming
galaxies around a quasar at $z=0.898$ using the TTF. We now use this
instrument to examine the fields of a small sample of six additional
quasars at $0.8 < z < 1.3$ for [O\,{\sc ii}] emission. Target field
selection and observing techniques are outlined in
\S\ref{sec:obs}. The detailed procedure for reducing these
observations is described in \S\ref{sec:red}, and results are
presented in \S\ref{sec:res}. Implications are discussed in
\S\ref{sec:dis}.

For a full discussion of the TTF and its use in finding emission-line
galaxies, the reader is encouraged to consult JBH01, Paper 1 or Jones,
Shopbell \& Bland-Hawthorn (2002; J02)\nocite{jones02}. The data
reduction presented here is as described in Paper 1, while the ELG
selection algorithm is that of J02. We adopt an $H_0 = 70$ km s$^{-1}$
Mpc$^{-1}$, $\Omega_{\Lambda} = 0.7$, flat-universe cosmology. Where
results of Paper 1 are incorporated, they are adjusted to these
values.

\section{Observations}
\label{sec:obs}

\subsection{Target selection}

Targets are drawn from the Molonglo Quasar Sample (MQS; Kapahi \etal
1998)\nocite{kapahi98} of low-frequency-selected RLQs. This is a
highly complete ($>97\%$) sample of 111 RLQs with $S_{408}>0.95$ Jy in
the declination range $-30^{\circ} < \delta < -20^{\circ}$, and
Galactic latitude $|b| > 20^{\circ}$ (excluding the R.A. ranges
\rashort{06}{00} -- \rashort{09}{00} and \rashort{14}{03} --
\rashort{20}{20}).  These quasars are, on average, five times less
powerful and so considerably more numerous at a given redshift 
than the rarer 3CR sources (Laing, Riley \& Longair 1983)\nocite{laing83}.

The six quasars whose fields are targetted in this work are listed in
Table~\ref{tab:tobs} and were chosen from the MQS parent sample purely
according to criteria of observability. Suitable redshifts are limited
by the need to place [O\,{\sc ii}] inside the wavelength ranges
accessible through the set of $\sim 200$\AA-wide TTF order-blocking
filters.  Additional constraints were solely due to RAs and
weather-affected runs, so the final list is a representative
selection. We emphasise that no preselection was invoked regarding the
likelihood of observing a cluster about a particular source.

The sample is chosen to be matched in radio power and therefore the
variance in this property is small, covering less than one decade. It
is therefore difficult to draw conclusions regarding the correlation
between the properties of the quasars or their environments, and radio
luminosity. We do not undertake any analysis of this type. It is
noted, however, that studies comparing clustering in the environments
of RLQs with their radio-quiet counterparts generally find that the
locales of each type of quasar are indistinguishable
\cite{wold00,wold01,finn01,mclure01}.

\begin{deluxetable*}{cccccccrcc}

  \tablewidth{0pt}
  \tablecolumns{10}
  \tablecaption{TTF targets and instrument
  parameters.\label{tab:tobs}}
  \tabletypesize{\scriptsize} 

  \tablehead{\colhead{Run} & \colhead{MRC} & \colhead{Obs.} &
   \colhead{$z$} & \colhead{$\lambda$([O\,{\sc ii}])} &
   \colhead{Blocking} & \colhead{Bandpass} & \colhead{Exposure} &
   \colhead{Seeing} & \colhead{Standard} \\
   \colhead{} & \colhead{quasar} & \colhead{Date} & \colhead{} &
   \colhead{} & \colhead{Filter} & \colhead{FWHM} & \colhead{time} &
   \colhead{} & \colhead{Star} \\
   \colhead{} & \colhead{} & \colhead{} & \colhead{} & \colhead{(\AA)}
   & \colhead{(\AA)} & \colhead{(\AA)} & \colhead{(s)} &
   \colhead{($^{\prime\prime}$)} & \colhead{} \\
   \colhead{(1)} & \colhead{(2)} & \colhead{(3)} & \colhead{(4)} &
   \colhead{(5)} & \colhead{(6)} & \colhead{(7)} & \colhead{(8)} &
   \colhead{(9)} & \colhead{(10)} \\
  }

      \startdata

        A & B0106--233 & 1999-Sep-07 & 0.818 & 6776 & 6680/210 & 10.5 &
        10500 & 1.7 & LTT 1020 \\
        B & B0413--210 & 1997-Oct-23 & 0.807 & 6735 & 6680/210 & 14.0
        & 7000 & 0.9 & LTT 1788 \\
        C & B1359--281 & 2000-Jul-29 & 0.802 & 6716 & 6680/210 & 10.2 &
        8400 & 1.4 & LTT 6248 \\
        D & B2021--208 & 2000-Jul-29 & 1.299 & 8568 & 8570/400 & 15.0 &
        16800 & 1.1 & LDS 749B \\
        E & B2037--234 & 1999-Sep-07 & 1.15\phn & 8013 & 8140/330 & 15.4
	& 14000 & 1.5 & LTT 7987 \\
        F & B2156--245 & 1997-Oct-23 & 0.862 & 6940 & 7070/260 & 11.6
        & 7000 & 0.9 & LTT 9239 \\

      \enddata

    \tablecomments{(1) run ID; (2) object name; (3) observation date;
    (4) redshift; (5) redshifted [O\,{\sc ii}]; (6) TTF blocking filter
    central wavelength/bandpass; (7) TTF scan bandpass; (8) total
    exposure time; (9) average seeing; (10) standard star used to
    calibrate the observation.}

\end{deluxetable*}

Paper 1 reported TTF observations of the field of 
one MQS quasar, MRC~B0450--221, the first RLQ observed
in our program to image ELGs in the environments of RLQs. The
reader is referred to this paper for details of the observations and
data reduction. Results from the examination of ELG candidates
in the field of MRC~B0450--221 will be incorporated into the
discussion of the collected properties of ELG candidates in the fields
of quasars (\S\ref{sec:dis}).

\subsection{Instrumental setup}

Observations of the MQS targets in this work were made using TTF at
f/8 on the AAT on 1997 October 23, 1999 September 7 and 2000 July
29. In each instance the MITLL2 CCD was windowed and masked to give a
\arcmn{9}{87} diameter circular field with a pixel scale of
\arcsd{0}{37}.

\begin{figure*}

  \epsscale{1}
  \plotone{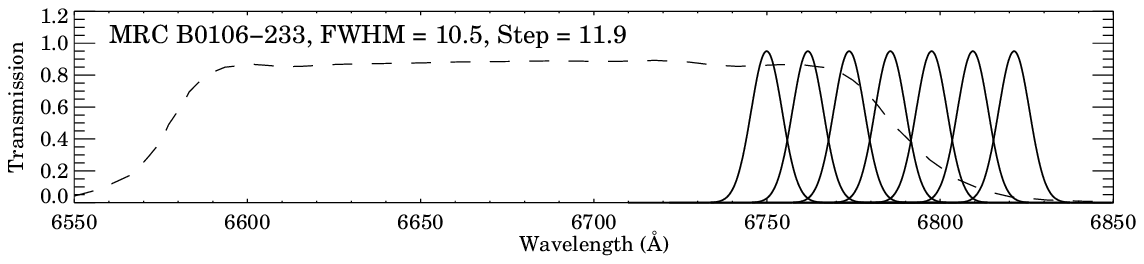}
  \plotone{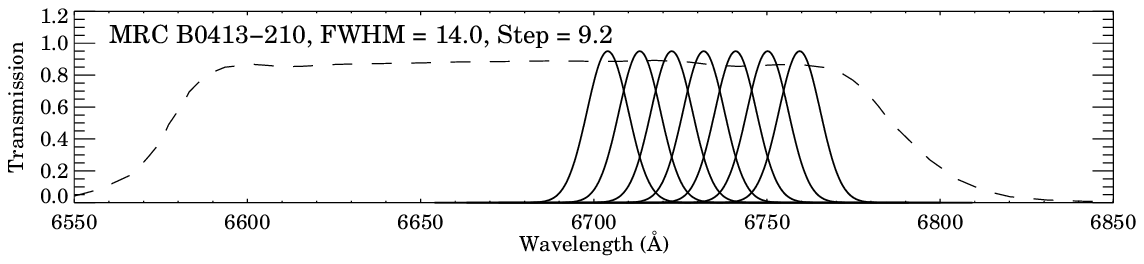}
  \plotone{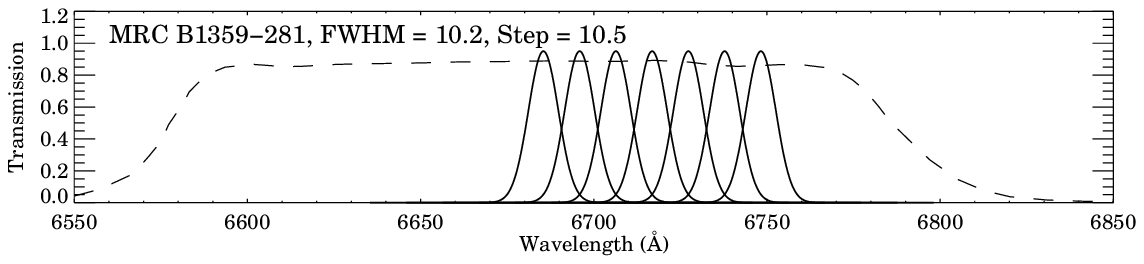}
  \plotone{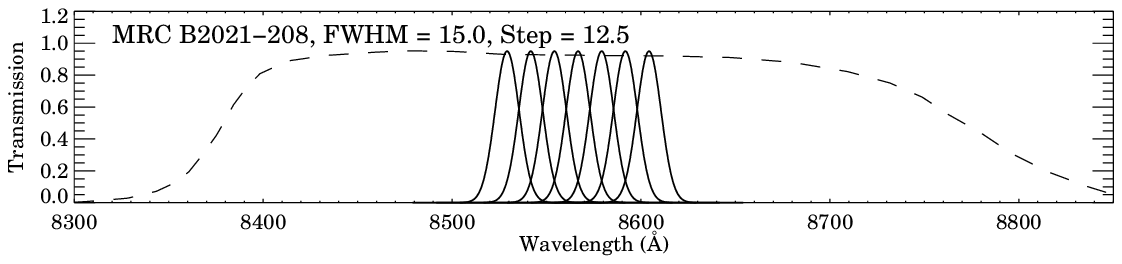}
  \plotone{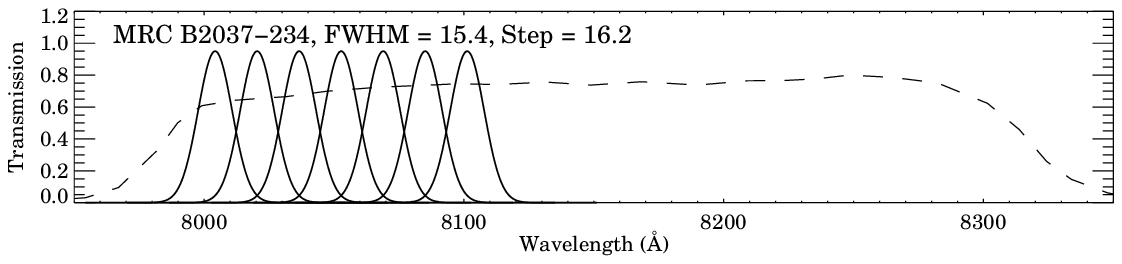}
  \plotone{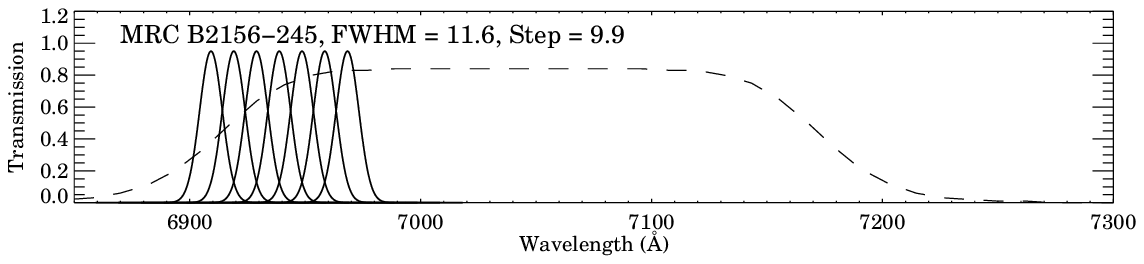}

  \figcaption{Passband sampling used in the TTF observations (profiles
  approximated by Gaussians: solid lines). The wavelengths shown are
  as measured at the location of the quasar near the field centre
  (note the wavelength sampled varies radially across the field for a
  given observation: see text). The transmission profiles of the
  order-blocking filters are shown as dashed lines. In the case of
  MRC~B2156--245, the $R1$ blocking filter has been tilted by
  $12^{\circ}$ in order to push its sensitivity $\sim 100$\AA \
  blueward. The FWHMs and steps are given in \AA.\label{fig:ban}}

\end{figure*}

TTF scanned the quasar fields at seven plate spacings, $Z$ ($\propto
\lambda$), corresponding to a series of steps of $\sim 10$\AA \ either
side of, and centred on, the redshifted [O\,{\sc ii}] line.  The transmission
profiles of the blocking filters and individual TTF scans are shown in
Figure~\ref{fig:ban} for each run. For run F, the blocking filter was
tilted by $12^{\circ}$ in order to push the sensitivity $\sim 100$\AA
\ blueward. At each plate spacing, $2 - 4$ images were taken in a
non-repeating pattern, with a relative spatial offset of $\sim
10^{\prime\prime}$ on the sky, to facilitate the removal of cosmic rays,
bad pixels and ghost images.

\begin{figure*}

  \epsscale{1}\plotone{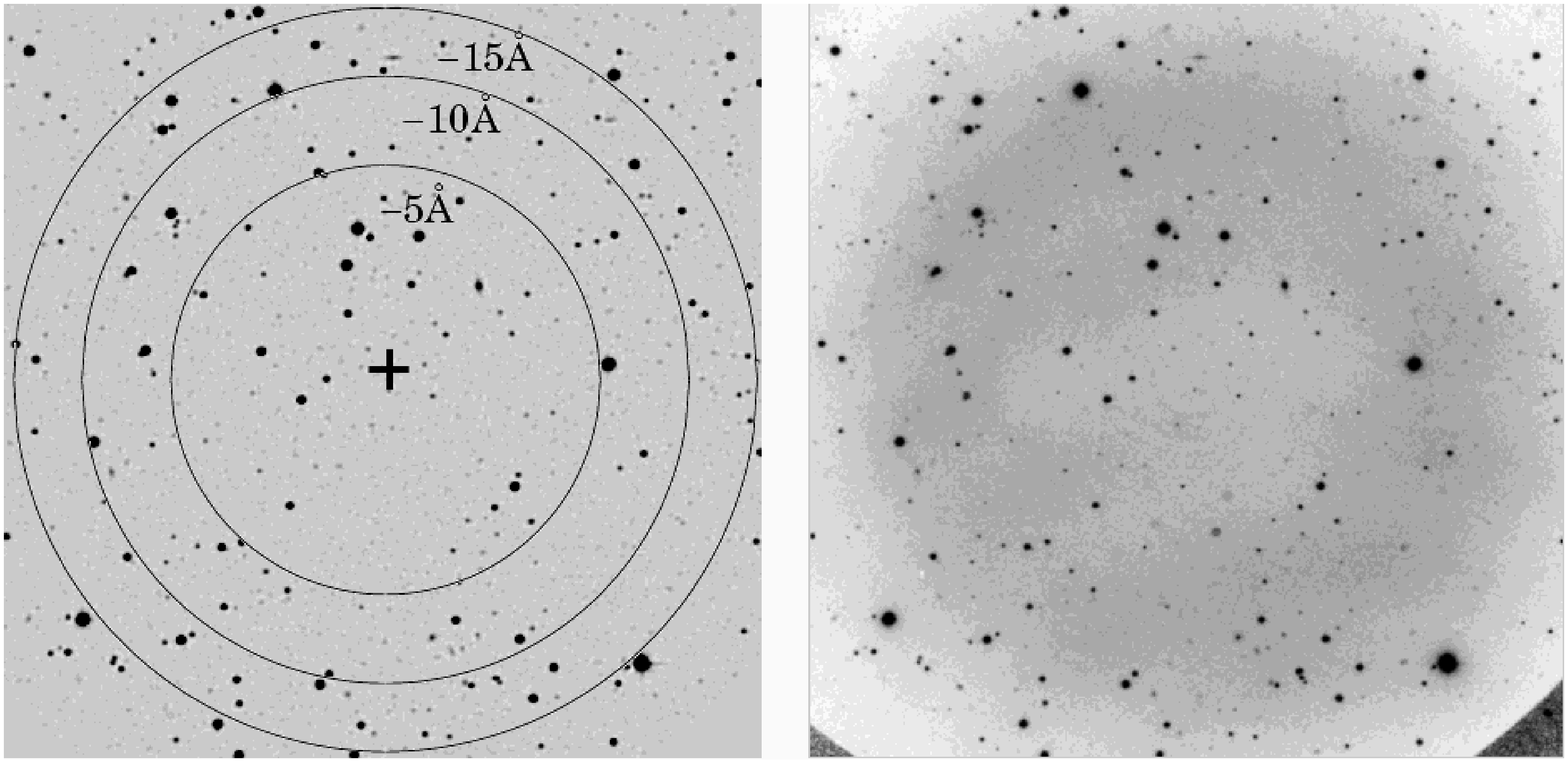}

  \figcaption{{\em Left:} Illustration of the wavelength variation
  across an example TTF field. The image is the central $7^{\prime}
  \times 7^{\prime}$ about MRC~B2021--208. Contours are drawn at 5\AA
  \ intervals from the optical axis which is marked with a '{\bf
  +}'. {\em Right:} A night-sky ring before removal from the same
  image.\label{fig:var}}

\end{figure*}

Fabry-Perot imagers have a quadratic radial wavelength dependence
($\lambda \propto r^2$); for the etalon tilts used in this study the
wavelength sensitivity varies across the field by 10 --
20\AA. Figure~\ref{fig:var} shows this variation for the field of
MRC~B2021--208, as well as a `night-sky ring', caused by the radial
sensitivity to emission from atmospheric OH. In order to measure this
spatial dependence, the wavelength scale was calibrated at two
positions; images of a CuAr or Ne lamp spectrum were made in 9 pixels
near the optical centre, where the wavelength sensitivity is at its
reddest, and in the same number at the field edge. This was
undertaken, over a range of $Z$ corresponding to $200-300$\AA, before
and after each data observation to mitigate any drift in the
$Z$,$\lambda$ relation. The $Z$,$\lambda$ and $r$,$\lambda$
relationships were thereby determined for each field and positions in
$x,y,Z$ space assigned the relevant wavelength solution. The change in
wavelength sensitivity with spatial position causes different volumes
to be sampled at different wavelengths. Because of the proximity of
the quasars to the optical centre in our images, more volume is
sampled blueward of the quasar redshift than redward. This is
accounted for when determining number densities in \S\ref{sec:dis}.

Flux calibrations were made by observing the spectrophotometric
standards indicated in Table \ref{tab:tobs}. For runs A, C, D and E,
exactly the same scan parameters were employed as were used in the
science observations. For scans B and F, the standards were observed
in a subset (2 or 3) of the $Z$-values used in the science frames. In
these latter cases the flux density was estimated assuming a linear
relationship between the counts in the standard at different
wavelengths.

\section{Data reduction}
\label{sec:red}

All data reduction and ELG candidate selection algorithms detailed
here were accomplished using IDL\footnote{Interactive data language;
http://www.rsinc.com/idl/} including custom scripts from the IDL
astronomy users
library\footnote{http://idlastro.gsfc.nasa.gov/homepage.html} as well
as programs written by the authors.

\subsection{Image production}

Data reduction initially proceeded as it would for standard broad-band
imaging. First the bias was removed from each image by subtracting the
median of a bias frame or the median value of the overscan region. The
data were flat-fielded at each $Z$ value using twilight-sky flats for
scans A, B, E and F and dome flats for scans C and D.

Night-sky rings, caused by radial sensitivity to OH Meinel bands (see
Figure~\ref{fig:var}), were removed by subtracting median values of
the sky in concentric circles about the optical axis. These rings are
actually ellipses, but for the etalon tilts and positions of the
optical centres in this study are well approximated by circles. The
night-sky rings were removed to within $4\%$ of the background level
in the most affected images.

Images at a common $Z$ value were aligned by comparing positions of
bright stars and co-added. Cosmic rays, bad pixels and ghosts were
rejected. The positions of the cosmic rays in $x,y,Z$ space were
recorded in order to cross-correlate their occurrence with that of
ELGs at a later stage.

\subsection{Catalogue creation}

Object detection and aperture photometry was carried out using
SExtractor (Bertin \& Arnouts)\nocite{bertin96}. Objects were detected
in two ways:

\begin{enumerate}

	\item SExtractor was used to identify objects at each $Z$
	value in the scan separately and these were correlated by
	position after extraction.

	\item The images from each scan were summed together and
	SExtractor was used to detect objects in this deep composite.

\end{enumerate}

\noindent
In both cases objects were identified as having 5 contiguous pixels at
$3\sigma$ above noise in a median-smoothed image convolved with the
scan's worst seeing. Each method has its advantages in terms of
locating ELGs. In the first case objects which are present at some $Z$
values but not others will not be diminished by the addition of extra
noise. In the second instance, fainter continuum objects will be
picked up, and the completeness limit of the catalogue will be
deeper. These two methods were used purely to identify the positions
at which to obtain aperture photometry in each image.

Photometry was then carried out using $3^{\prime\prime}$ diameter
apertures. The objects were calibrated to a flux scale (erg s$^{-1}$
cm$^{-2}$) and AB magnitudes using the spectrophotometric standards
indicated in Table~\ref{tab:tobs}. A small offset of typically 0.1 mag
was applied to each object to correct for flux from the object falling
outside the fixed aperture. This was calculated by comparing the
magnitudes of a subset of $\sim 100$ objects measured within the
$3^{\prime\prime}$ apertures with the magnitudes measured in larger
($5^{\prime\prime} - 7^{\prime\prime}$) apertures. Corrections for
galactic extinction (typically $A_{V} \sim 0.1$ mag) were made to each
field using values from the NASA/IPAC Extragalactic
Database\footnote{http://nedwww.ipac.caltech.edu/}.

The two catalogues, compiled from objects in individual bands and
those identified from the combined scan, were analysed separately for
the presence of candidate emission-line galaxies.

\section{Identification of ELGs in TTF stacks}
\label{sec:res}

Objects detected by SExtractor at only one, two or three adjacent etalon
values in the individual catalogues were identified first and set
aside as ELG candidates. This initial sweep separated those objects
for which only line emission was detectable. The search then proceeded
to objects with line emission superposed on a continuum.

For objects detected individually in four or more images, a straight
line was fitted to their fluxes in each band. The rms scatter,
$\sigma$, about the line and the mean flux error, $\langle \Delta F
\rangle$, were evaluated and the larger taken as the dominant source
of error, $\sigma_{\mathrm{dom}}$. The background was then fitted
iteratively as a straight line, rejecting the points that fell more
than $\sigma_{\mathrm{dom}}$ above the line. A minimum of three points
were retained to fit the background. Objects with peaks $>3\sigma$
above the fitted background were set aside as candidate line
emitters. Those $Z$ values in which the flux was $>3\sigma$ above the
background were identified as `line' bands under two conditions:

\begin{enumerate}

\item Where there are two or more bands above the threshold, they
should be adjacent.

\item If only one band is above the threshold, the (one or) two images
adjacent to it are classified as `line' bands. This is because line
emission superposed on a detectable continuum is never narrow enough
to be contained only in one $\sim 10$\AA \ wavelength slice.

\end{enumerate}

\noindent

\begin{figure*}

  \epsscale{1}
  \plottwo{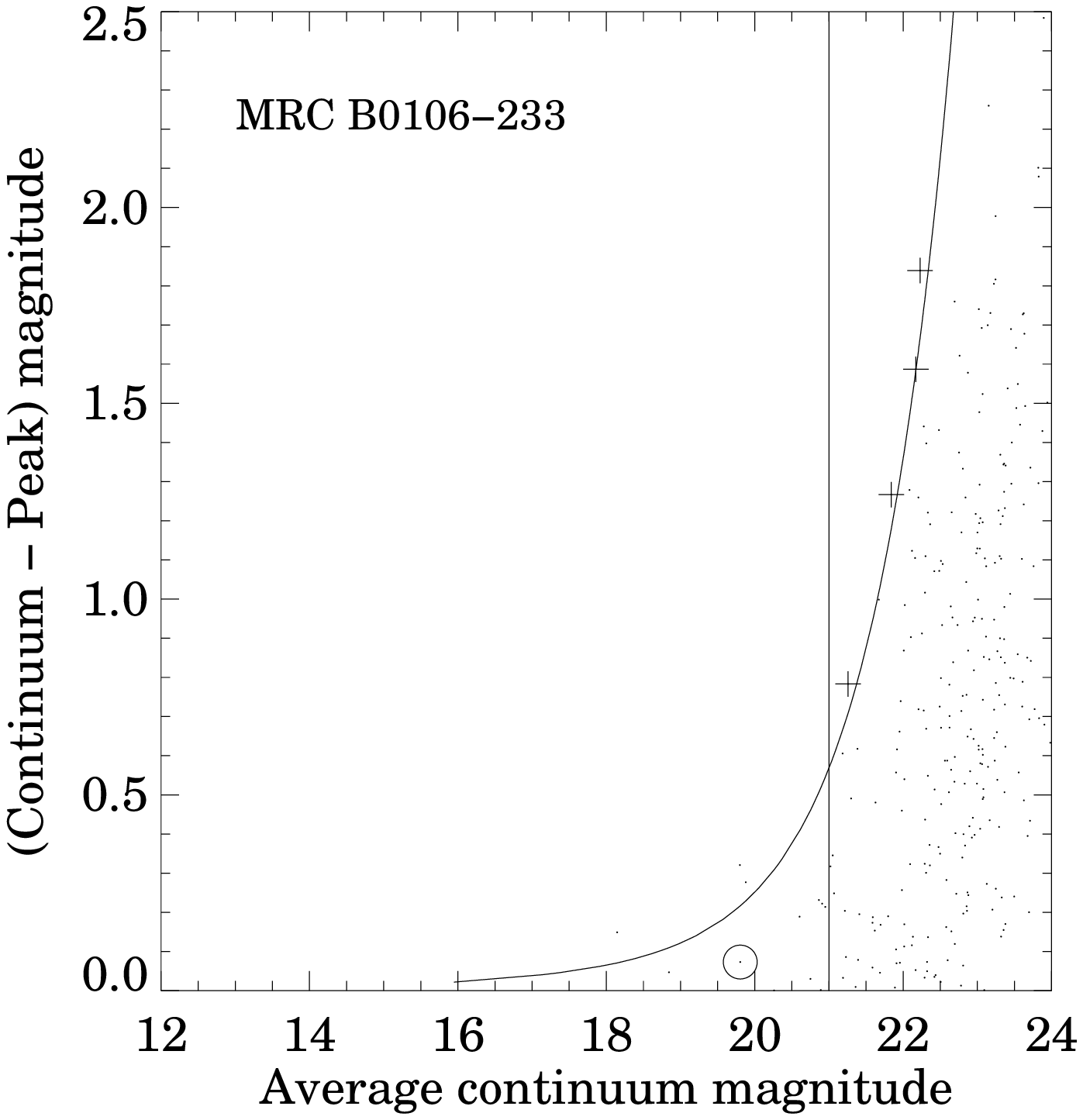}{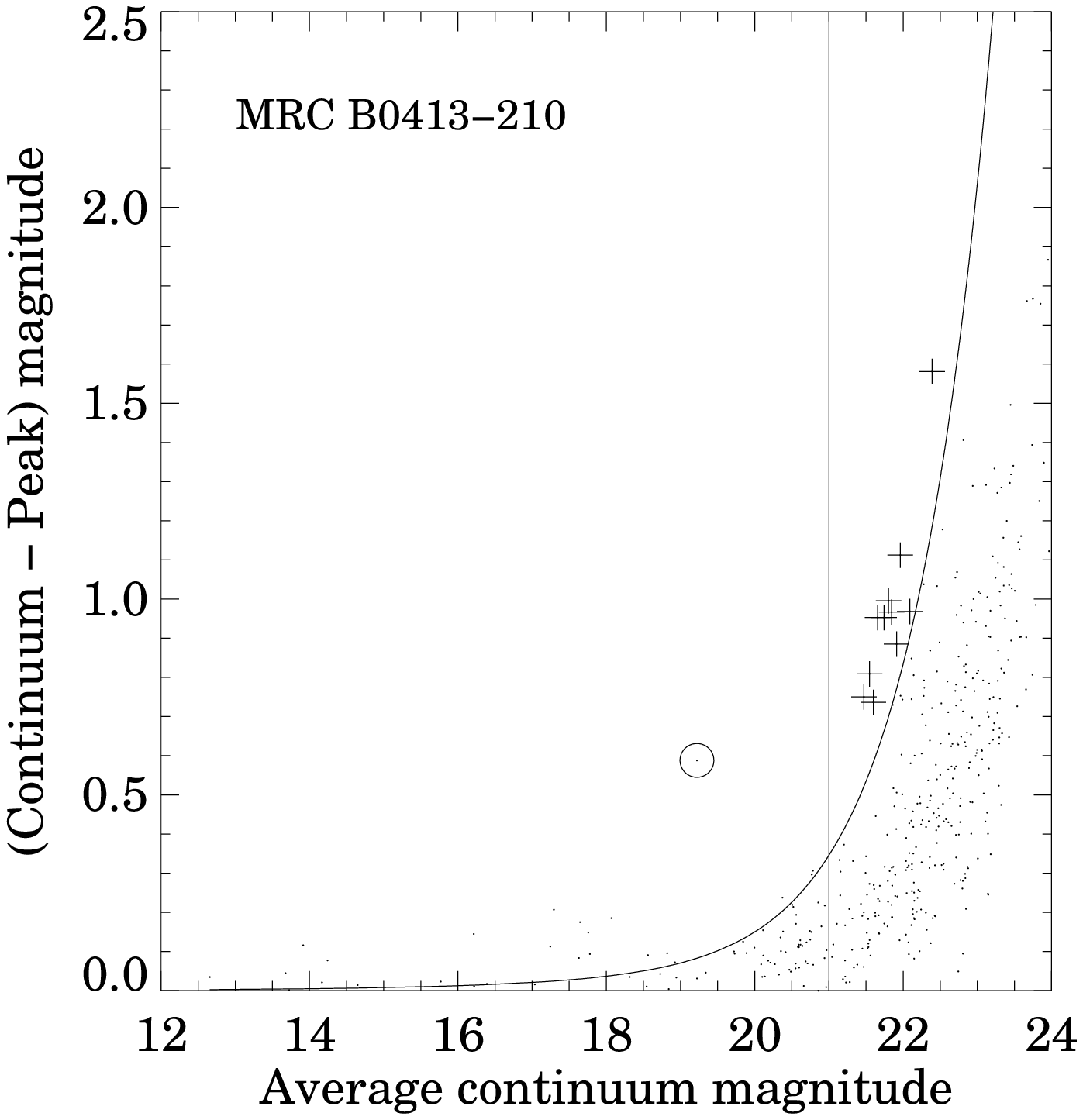}
  \plottwo{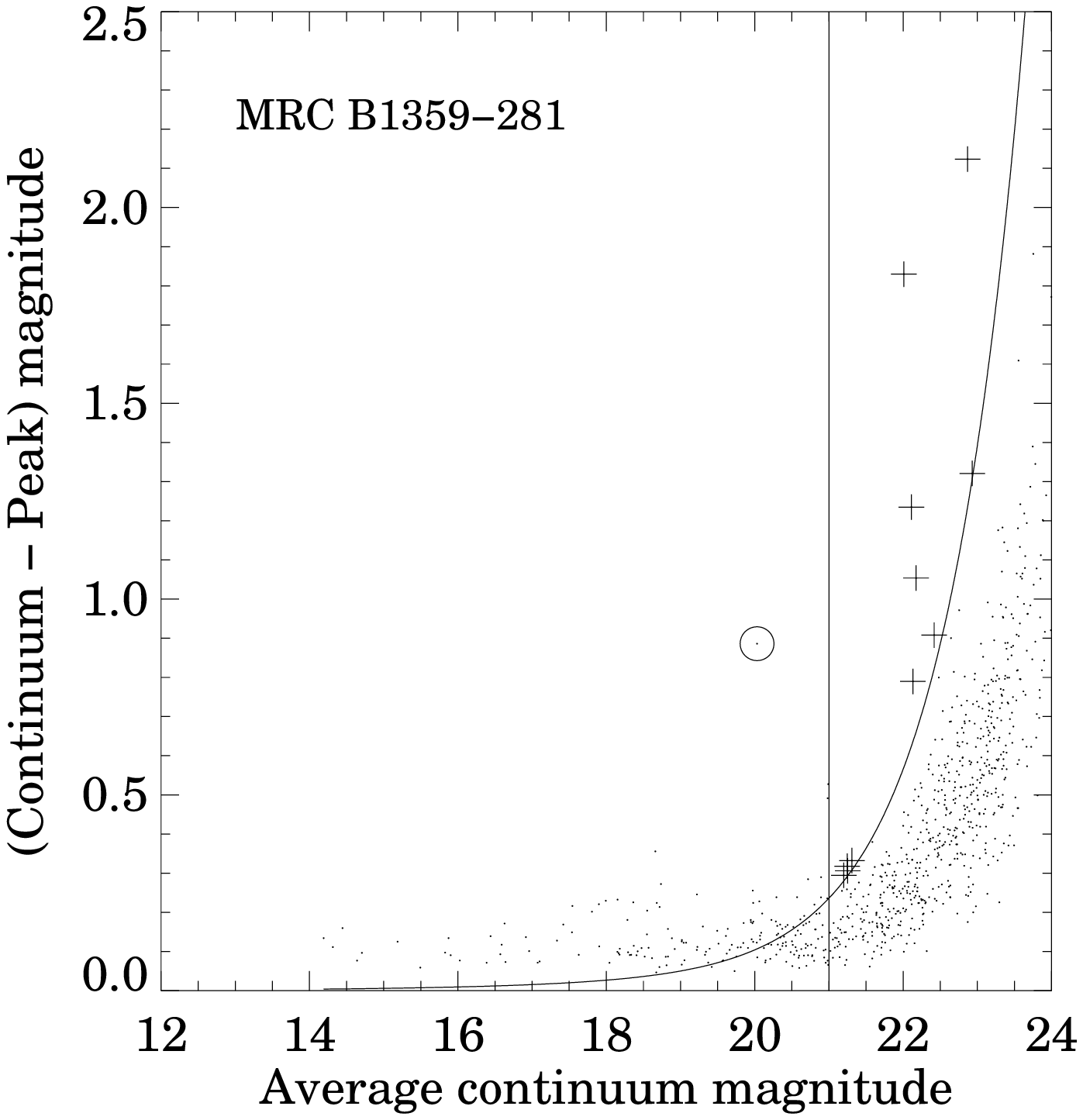}{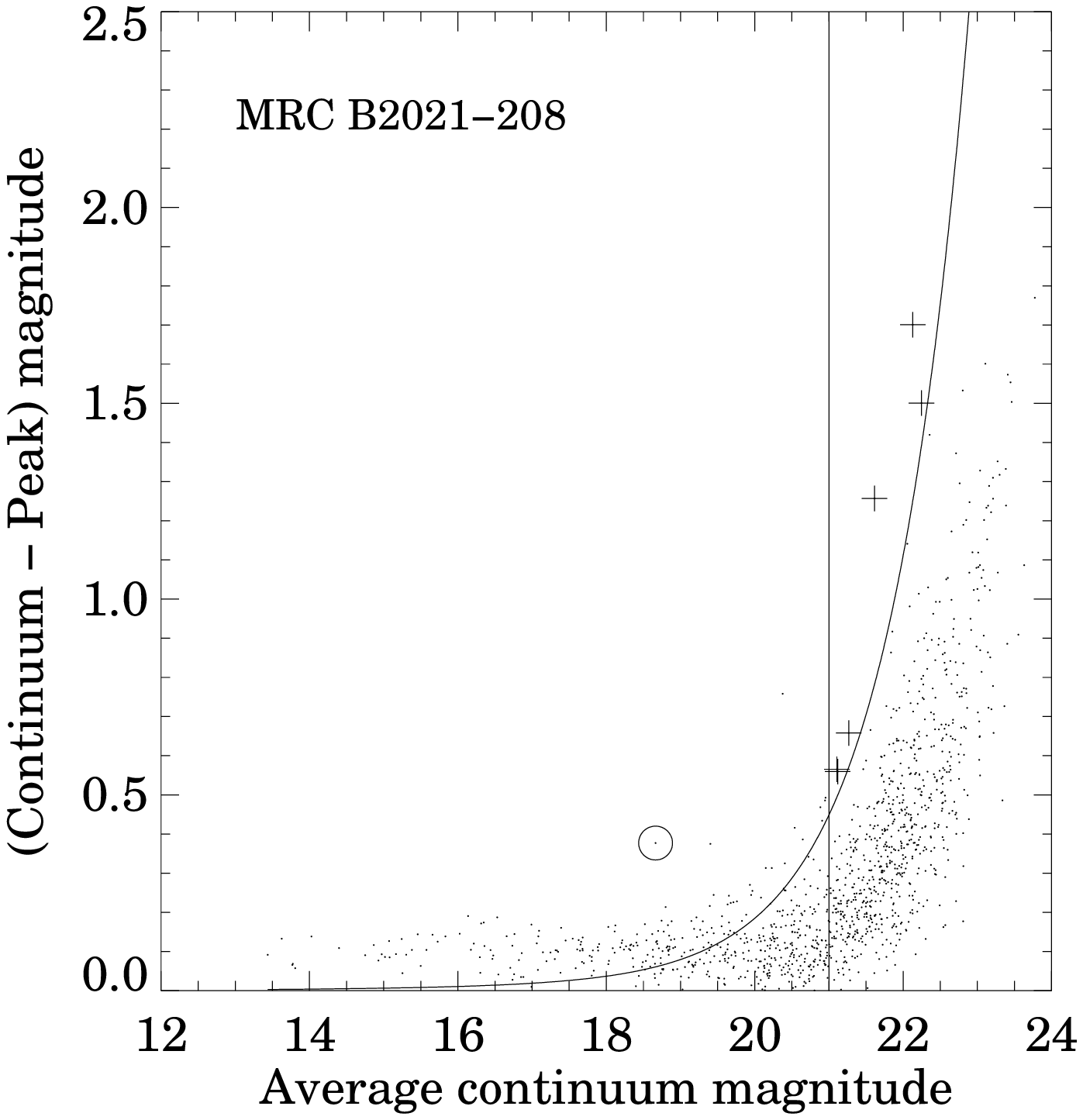}
  \plottwo{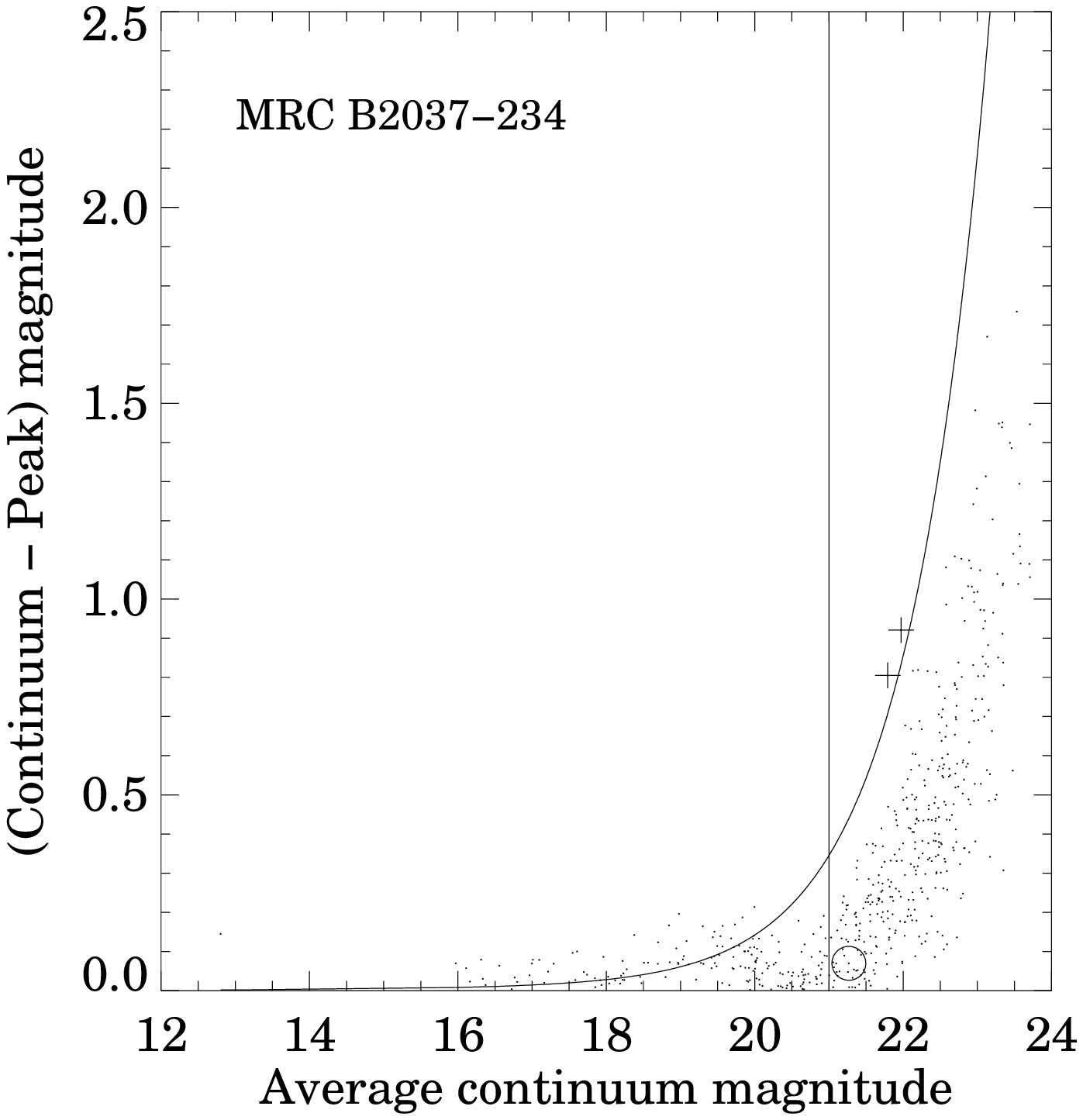}{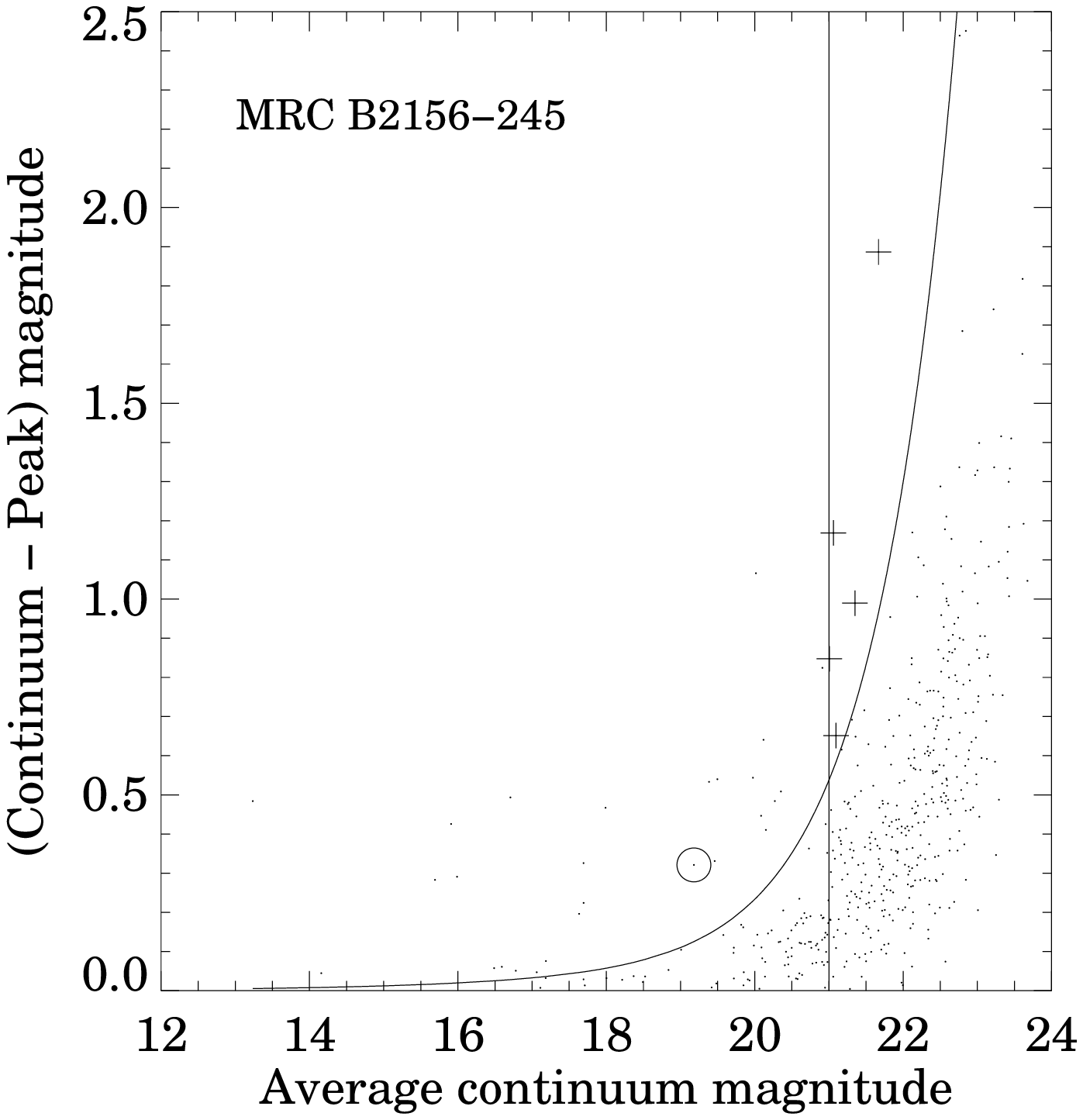}

  \figcaption[Barr.fig3a.eps,Barr.fig3b.eps,Barr.fig3c.eps,Barr.fig3d.eps,Barr.fig3e.eps,Barr.fig3f.eps]{Difference
  between estimated peak and continuum magnitude (measured in all
  seven frames) plotted as a function of continuum magnitude for
  objects in the field of all quasars. The solid lines show the ELG
  selection criteria, \ie $I$(AB) $>21$ and continuum $-$ peak mag
  $=3\times$ the error in the average continuum magnitude. Objects
  qualifying as ELG candidates are marked as crosses. The quasar is
  marked by a circle.\label{fig:m_p}}

\end{figure*}

Candidate ELGs were also identified by considering the difference
between the line and continuum magnitudes. ELG candidates are
identified as objects with brightest magnitudes separated by more than
three times the error from the average continuum magnitude. The
selection is illustrated in Figure~\ref{fig:m_p} for each field. A
limit of $I(AB) > 21$ is imposed on putative line emitters to avoid
contamination of the sample at the bright end, where differences in
luminosity between bands caused by light falling outside the small
aperture become significant. At the redshifts probed by our sample we
do not expect ELGs with $I(AB) > 21$. This method was used
successfully to find ELGs in the field of MRC~B0450--221 in Paper 1
and in Hall \etal (2001\nocite{hall01}).

\begin{figure*}

  \epsscale{1}
  \plotone{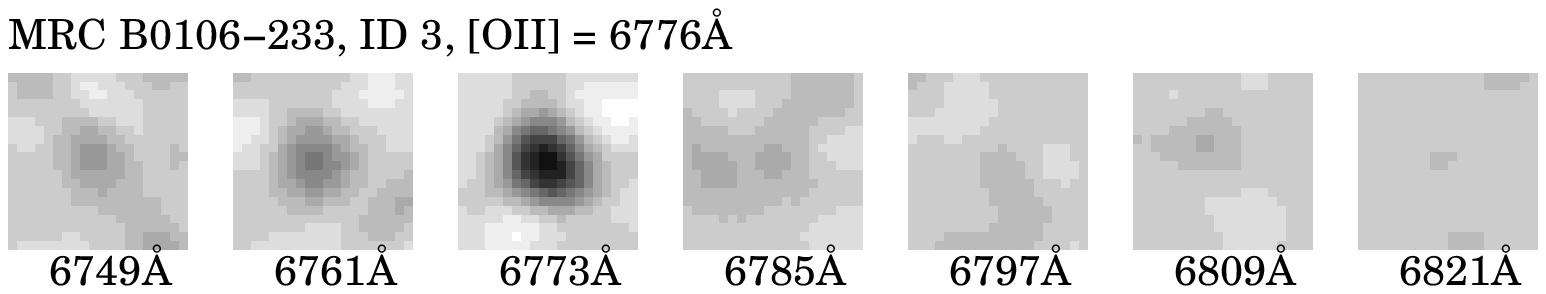}
  \plotone{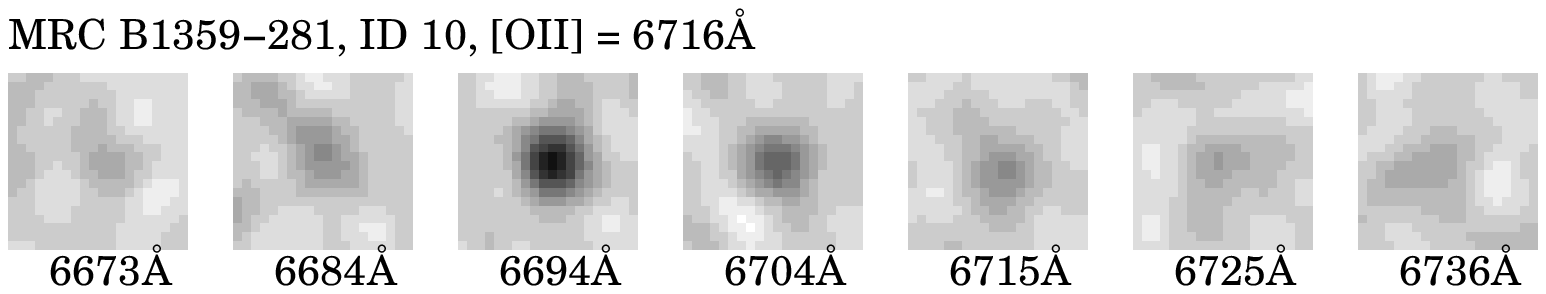}
  \plotone{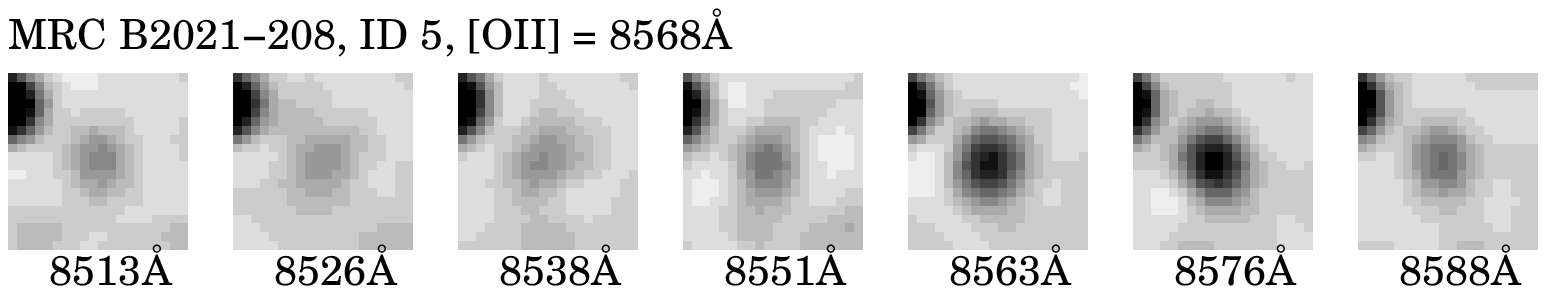}
  \plotone{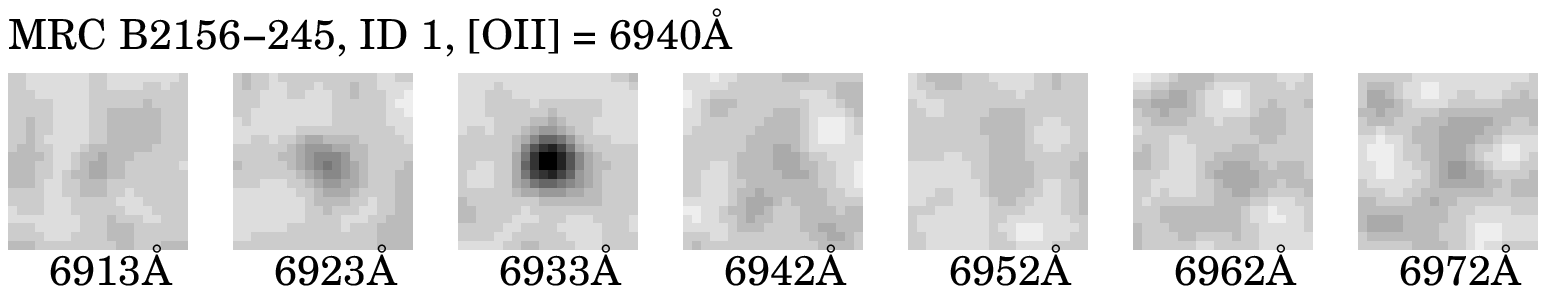}
  \plotone{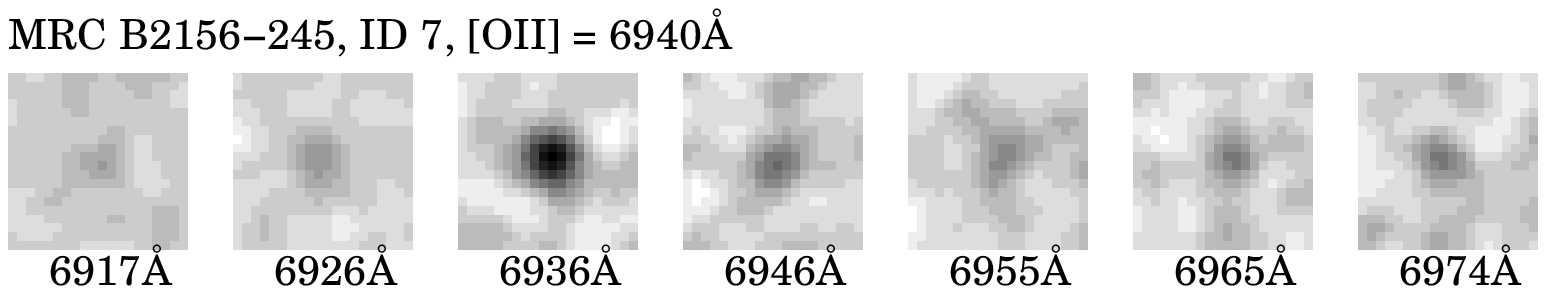}
  \plotone{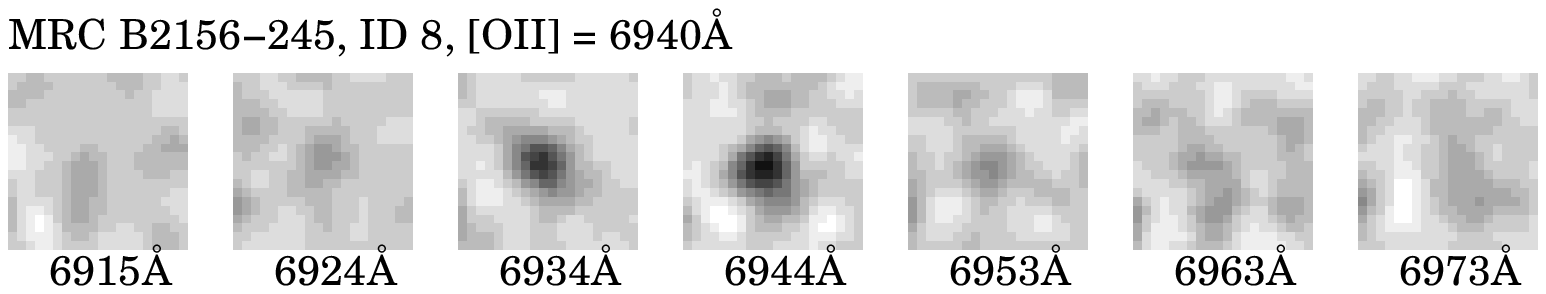}

  \figcaption[Barr.fig4a.eps,Barr.fig4b.eps,Barr.fig4c.eps,Barr.fig4d.eps,Barr.fig4e.eps,Barr.fig4f.eps]{Images
  showing some candidate line emitters in the fields of quasars. Each
  row corresponds to an object (with quasar name, ELG candidate ID and
  wavelength of redshifted [O\,{\sc ii}] labelled top-left) at
  different wavelengths. Each frame is \arcsd{7}{5} square and is
  labelled with its central wavelength. North is up, East is
  left.\label{fig:elg}}

\end{figure*}

All ELG candidates were cross-checked in $x,y,Z$ space with those of
cosmic rays, and matches were excluded from the catalogue. Generally,
the same objects were extracted by each selection method. However, the
overlap was not complete -- an indication that some candidates are
missed by each process (see \S\ref{sec:elginc}). As a final check, all
candidates were examined by eye to ensure that no unusual cosmic rays
or low-level ghosts were categorised as ELGs. A selection of
candidates are shown in Figure~\ref{fig:elg}.

A total of 47 objects were identified as ELG candidates in our TTF
observations. At least five candidates were discovered in each
field. The characteristics of the ELG candidates found using TTF are
shown in Table~\ref{tab:elg_1}. Broad-band $I$ magnitudes are listed
for those objects that have been detected as part of a separate
program to detect clustering in the fields of RLQs
\cite{barr03t,barr03}. Redshifts are assigned based on the wavelength
bin in which the emission peaks. The errors are assigned,
conservatively, as the entire width of the wavelength bin. For those
quasars for which [O\,{\sc ii}] emission was detected, MRC~B0413--210,
MRC~B1359--281 and MRC~B2156--245, the redshift was checked for
consistency with the published spectrum. In the latter two cases the
peak in the [O\,{\sc ii}] emission agrees with the spectra published
in Baker \etal (1999)\nocite{baker99}. No occurrence of [O\,{\sc ii}]
is documented for MRC~B0413--210 in Baker \etal For this quasar the
wavelength calibration is not secure and is rederived assuming that
[O\,{\sc ii}] peaks at the quasar redshift (see \S\ref{sec:0413}). The
[O\,{\sc ii}] fluxes for MRC~B1359--281 and MRC~B2156--245 in Baker
\etal (1999) were also found to be consistent, within the errors, with
the present work.

\begin{deluxetable*}{cccccc@{$\pm$}crrr@{$\pm$}lrr@{$\pm$}l}
\renewcommand{\arraystretch}{.6}

  \tablecaption{Positions, redshift and luminosity characteristics of
  ELG candidates\label{tab:elg_1}}
  \tablewidth{0pt}
  \tabletypesize{\scriptsize}
  \tablecolumns{14}

  \tablehead{\colhead{MRC} & \colhead{ID} &
   \multicolumn{2}{c}{Position} & \colhead{$\lambda_{p}$} &
   \multicolumn{2}{c}{$z$([O\,{\sc ii}])} &
   \multicolumn{2}{c}{Continuum} & \multicolumn{2}{c}{$F_{l}$} &
   \colhead{$W_{\lambda}$} & \multicolumn{2}{c}{$L$([O\,{\sc ii}])}
   \\
   \colhead{quasar} & \colhead{} & \multicolumn{2}{c}{(J2000.0)} &
   \colhead{} & \colhead{} & \colhead{} &
   \multicolumn{2}{c}{magnitude} & \multicolumn{2}{c}{($\times
   10^{-16}$ erg} & \colhead{} & \multicolumn{2}{c}{($ \times
   10^{41}$} \\
   \colhead{} & \colhead{} & \colhead{R.A.} & \colhead{Decl.} &
   \colhead{(\AA)} & \colhead{} & \colhead{} & \colhead{$I(AB)$} &
   \colhead{$I$} & \multicolumn{2}{c}{s$^{-1}$ cm$^{-2}$)} &
   \colhead{(\AA)} & \multicolumn{2}{c}{erg s$^{-1}$)} \\
   \colhead{(1)} & \colhead{(2)} & \colhead{} & \colhead{} & \colhead{(3)} &
   \multicolumn{2}{c}{(4)} & \colhead{(5)} & \colhead{(6)} &
   \multicolumn{2}{c}{(7)} & \colhead{(8)} & \multicolumn{2}{c}{(9)} \\
  }

  \startdata

        B0106--233 \\ 

 & $ 1$ & $01$ $09$ $ \: \;  3.08$ & $-23$ $05$ $19.81$ & $6754$ & $0.812$ & $0.003$ & $>23.5$ & $>21.0$ & $  0.3$ & $  0.0$ & $> 24.2$ & $  0.8$ & $  0.1$ \\
 & $ 2$ & $01$ $09$ $ \: \;  4.10$ & $-23$ $07$ $16.26$ & $6773$ & $0.817$ & $0.003$ & $>23.5$ & 21.8 & $  0.4$ & $  0.0$ & $>  15.8$ & $  1.2$ & $  0.1$ \\
 & $ 3$ & $01$ $09$ $ \: \;  5.00$ & $-23$ $07$ $27.67$ & $6773$ & $0.817$ & $0.003$ & $>23.5$ & 22.7 & $  1.3$ & $  0.1$ & $> 93.5$ & $ 3.8$ & $  0.3$ \\
 & $ 4$ & $01$ $09$ $ \: \;  0.95$ & $-23$ $05$ $34.20$ & $6767$ & $0.816$ & $0.003$ & $>23.5$ & 22.4 & $  0.6$ & $  0.1$ & $> 56.7$ & $ 1.9$ & $  0.3$ \\
 & $ 5$ & $01$ $08$ $46.25$ & $-23$ $06$ $23.55$ & $6782$ & $0.820$ & $0.003$ & $>23.5$ & 22.4 & $  0.2$ & $  0.1$ & $>  11.6$ & $  0.6$ & $  0.2$ \\

       B0413--210 \\

 & $ 1$ & $04$ $16$ $ \: \;  8.85$ & $-20$ $58$ $ \: \;  9.33$ & $6730$ & $0.806$ & $0.002$ & $>23.0$ & \nodata & $  0.5$ & $  0.2$ & $>  28.0$ & $  1.5$ & $  0.5$ \\
 & $ 2$ & $04$ $15$ $54.14$ & $-20$ $58$ $ \: \;  2.83$ & $6731$ & $0.806$ & $0.002$ & $>23.0$ & \nodata & $  0.4$ & $  0.2$ & $>  22.4$ & $  1.2$ & $  0.5$ \\
 & $ 3$ & $04$ $16$ $ \: \;  5.16$ & $-20$ $56$ $43.85$ & $6738$ & $0.808$ & $0.002$ & $>23.0$ & \nodata & $  1.0$ & $  0.2$ & $> 50.4$ & $ 2.9$ & $  0.5$ \\
 & $ 4$ & $04$ $16$ $ \: \;  7.03$ & $-20$ $57$ $41.73$ & $6730$ & $0.806$ & $0.002$  & $>23.0$ & \nodata & $  0.2$ & $  0.2$ & $>  9.8$ & $  0.6$ & $  0.5$ \\
 & $ 5$ & $04$ $15$ $58.06$ & $-20$ $52$ $33.64$ & $6719$ & $0.803$ & $0.002$ & $>23.0$ & \nodata & $  0.6$ & $  0.2$ & $> 29.4$ & $ 1.8$ & $  0.5$ \\
 & $ 6$ & $04$ $16$ $ \: \;  4.08$ & $-20$ $56$ $15.48$ & $6737$ & $0.808$ & $0.002$ & $>23.0$ & \nodata & $  0.6$ & $  0.2$ & $> 30.8$ & $  1.8$ & $  0.5$ \\
 & $ 7$ & $04$ $15$ $56.70$ & $-20$ $54$ $11.00$ & $6738$ & $0.808$ & $0.002$ & $>23.0$ & \nodata & $  0.4$ & $  0.2$ & $>  18.2$ & $  1.1$ & $  0.5$ \\
 & $ 8$ & $04$ $15$ $52.58$ & $-20$ $58$ $42.42$ & $6731$ & $0.806$ & $0.002$ & $>23.0$ & \nodata & $  0.9$ & $  0.2$ & $> 56.0$ & $ 2.6$ & $  0.5$ \\
 & $ 9$ & $04$ $16$ $14.41$ & $-20$ $57$ $48.46$ & $6717$ & $0.802$ & $0.002$ & $>23.0$ & \nodata & $  1.5$ & $  0.2$ & $> 182.0$ & $ 4.5$ & $  0.5$ \\
 & $ 10$ & $04$ $16$ $ \: \;  2.51$ & $-21$ $00$ $23.73$ & $6741$ & $0.809$ & $0.002$ & $>23.0$ & \nodata & $  0.3$ & $  0.1$ & $> 11.2$ & $ 0.9$ & $  0.4$ \\
 & $ 11$ & $04$ $16$ $17.77$ & $-20$ $56$ $54.49$ & $6740$ & $0.808$ & $0.002$ &  22.2 & \nodata & $  1.0$ & $  0.2$ &   28.0 & $ 2.8$ & $  0.7$ \\

        B1359--281 \\

 & $ 1$ & $14$ $01$ $54.79$ & $-28$ $21$ $17.70$ & $6699$ & $0.797$ & $0.002$ & $>23.0$ & 21.5  & $  0.8$ & $  0.1$ & $> 128.5$ & $ 2.2$ & $  0.4$ \\
 & $ 2$ & $14$ $02$ $ \: \;  4.48$ & $-28$ $23$ $ \: \;  6.59$ & $6717$ & $0.802$ & $0.002$  & $>23.0$ & 21.6 & $  1.1$ & $  0.1$ & $> 186.7$ & $ 3.2$ & $  0.4$ \\
 & $ 3$ & $14$ $01$ $51.94$ & $-28$ $25$ $ \: \;  2.49$ & $6702$ & $0.798$ & $0.002$ & $>23.0$ & \nodata & $  0.8$ & $  0.1$ & $> 130.6$ & $ 2.2$ & $  0.4$ \\
 & $ 4$ & $14$ $02$ $22.02$ & $-28$ $21$ $30.83$ & $6711$ & $0.801$ & $0.002$  & $>23.0$ & \nodata & $  1.1$ & $  0.1$ & $> 187.7$ & $ 3.2$ & $  0.4$ \\
 & $ 5$ & $14$ $02$ $13.67$ & $-28$ $20$ $23.29$ & $6707$ & $0.800$ & $0.002$  & $>23.0$ & 22.5 & $  1.0$ & $  0.1$ & $> 159.1$ & $ 2.7$ & $  0.4$ \\
 & $ 6$ & $14$ $01$ $46.25$ & $-28$ $22$ $28.22$ & $6719$ & $0.803$ & $0.002$ & $>23.0$ & \nodata & $  0.7$ & $  0.1$ & $> 121.4$ & $ 2.0$ & $  0.4$ \\
 & $ 7$ & $14$ $02$ $10.07$ & $-28$ $18$ $ \: \;  7.38$ & $6713$ & $0.801$ & $0.002$ &  21.2 & \nodata & $ 1.2$ & $  0.2$ &    10.2 & $ 3.4$ & $  0.5$ \\
 & $ 8$ & $14$ $02$ $ \: \;  9.51$ & $-28$ $18$ $ \: \;  1.09$ & $6702$ & $0.798$ & $0.002$ &  22.4 & \nodata & $  0.6$ & $  0.1$ &   21.4 & $ 1.8$ & $  0.4$ \\
 & $ 9$ & $14$ $01$ $55.77$ & $-28$ $18$ $41.35$ & $6704$ & $0.799$ & $0.002$ &  22.0 & \nodata & $  1.3$ & $  0.1$ &   101.0 & $ 3.7$ & $  0.4$ \\
 & $ 10$ & $14$ $01$ $53.33$ & $-28$ $20$ $34.70$ & $6694$ & $0.796$ & $0.002$ &  22.1 & 21.9 & $  1.7$ & $  0.1$ &   56.1 & $ 4.8$ & $  0.4$ \\

        B2021--208 \\

 & $ 1$ & $20$ $24$ $38.52$ & $-20$ $42$ $32.19$ & $8525$ & $1.287$ & $0.003$ & $>22.0$ & \nodata & $  0.7$ & $  0.1$ & $> 51.0$ & $ 6.6$ & $  1.1$ \\
 & $ 2$ & $20$ $24$ $44.03$ & $-20$ $46$ $10.44$ & $8544$ & $1.292$ & $0.003$ & $>22.0$ & \nodata & $  1.1$ & $  0.1$ & $> 85.5$ & $ 10.0$ & $  1.1$ \\
 & $ 3$ & $20$ $24$ $39.45$ & $-20$ $44$ $31.53$ & $8528$ & $1.288$ & $0.003$ &  21.3 & \nodata & $  0.6$ & $  0.1$ &    12.0 & $ 5.6$ & $  1.1$ \\
 & $ 4$ & $20$ $24$ $41.10$ & $-20$ $40$ $19.61$ & $8556$ & $1.296$ & $0.003$ &  21.6 & \nodata & $  0.2$ & $  0.2$ &    4.5 & $ 2.0$ & $  1.6$ \\
 & $ 5$ & $20$ $24$ $41.63$ & $-20$ $41$ $ \: \;  5.41$ & $8576$ & $1.301$ & $0.003$ &  20.8 & \nodata & $ 2.9$ & $  0.2$ &   40.5 & $27.7$ & $  1.4$ \\
 & $ 6$ & $20$ $24$ $46.52$ & $-20$ $41$ $ \: \;  7.71$ & $8520$ & $1.286$ & $0.003$ &  20.6 & \nodata & $  1.3$ & $  0.1$ &   36.0 & $ 12.0$ & $  1.1$ \\
 & $ 7$ & $20$ $24$ $52.36$ & $-20$ $41$ $44.82$ & $8525$ & $1.287$ & $0.003$  &  21.2 & \nodata & $  1.0$ & $  0.1$ &   28.5 & $ 9.3$ & $  1.3$ \\

        B2037--234 \\

 & $ 1$ & $20$ $40$ $ \: \;  8.42$ & $-23$ $17$ $37.70$ & $8053$ & $1.161$ & $0.004$ & $>23.0$ & 22.0 & $  1.0$ & $  0.2$ & $> 55.0$ & $ 7.1$ & $  1.4$ \\
 & $ 2$ & $20$ $40$ $13.85$ & $-23$ $16$ $40.02$ & $8047$ & $1.159$ & $0.004$ & $>23.0$ & $>22.0$ & $  0.4$ & $  0.2$ & $> 24.6$ & $ 3.1$ & $  1.3$ \\
 & $ 3$ & $20$ $39$ $55.87$ & $-23$ $14$ $38.31$ & $8036$ & $1.156$ & $0.004$ & $>23.0$ & 22.8 & $  1.0$ & $  0.2$ & $> 41.6$ & $ 6.8$ & $  1.5$ \\
 & $ 4$ & $20$ $39$ $57.91$ & $-23$ $18$ $ \: \;  3.52$ & $8072$ & $1.166$ & $0.004$ & $>23.0$ & 21.9 & $  1.1$ & $  0.2$ & $> 37.0$ & $ 7.5$ & $  1.1$ \\
 & $ 5$ & $20$ $40$ $ \: \;  2.37$ & $-23$ $15$ $ \: \;  6.76$ & $8056$ & $1.162$ & $0.004$ & 21.8 & 21.5 & $  0.5$ & $  0.2$ &   23.1 & $ 3.4$ & $  1.1$ \\

	B2156--245 \\

 & $ 1$ & $21$ $59$ $27.74$ & $-24$ $19$ $ \: \;  5.61$ & $6933$ & $0.860$ & $0.002$ & $>22.5$ & 21.6 & $ 4.0$ & $  0.3$ & $> 129.9$ & $ 13.7$ & $  0.9$ \\
 & $ 2$ & $21$ $59$ $21.06$ & $-24$ $14$ $11.40$ & $6898$ & $0.851$ & $0.003$ & $>22.5$ & 21.3 & $  1.4$ & $  0.3$ & $> 36.0$ & $ 4.6$ & $  1.0$ \\
 & $ 3$ & $21$ $59$ $36.87$ & $-24$ $20$ $14.96$ & $6955$ & $0.866$ & $0.002$ & $>22.5$ & 21.3 & $  1.1$ & $  0.2$ & $> 44.1$ & $ 3.7$ & $  0.7$ \\
 & $ 4$ & $21$ $59$ $17.67$ & $-24$ $20$ $10.58$ & $6965$ & $0.869$ & $0.002$ & $>22.5$ & 21.4 & $  0.4$ & $  0.2$ & $>  17.4$ & $  1.3$ & $  0.5$ \\
 & $ 5$ & $21$ $59$ $42.20$ & $-24$ $20$ $ \: \;  5.31$ & $6946$ & $0.864$ & $0.002$ &$ >22.5$ & 21.3 & $  1.4$ & $  0.2$ & $> 49.9$ & $ 4.9$ & $  0.8$ \\
 & $ 6$ & $21$ $59$ $16.05$ & $-24$ $17$ $ \: \;  5.88$ & $6944$ & $0.863$ & $0.002$ &  21.3 & 21.1 & $  0.9$ & $  0.2$ &   24.4 & $ 3.0$ & $  0.7$ \\
 & $ 7$ & $21$ $59$ $18.30$ & $-24$ $19$ $37.35$ & $6936$ & $0.861$ & $0.002$ & 21.0 & 21.1 & $ 3.6$ & $  0.3$ &   85.8 & $12.3$ & $  1.0$ \\
 & $ 8$ & $21$ $59$ $21.12$ & $-24$ $18$ $48.38$ & $6944$ & $0.863$ & $0.002$ &  20.4 & 21.3 & $ 3.3$ & $  0.2$ &   65.0 & $ 11.1$ & $  0.8$ \\
 & $ 9$ & $21$ $59$ $25.45$ & $-24$ $17$ $45.32$ & $6948$ & $0.864$ & $0.002$ &  20.9 & 20.7 & $ 2.4$ & $  0.3$ &   53.4 & $ 8.1$ & $  0.9$ \\

      \enddata

	\tablecomments{(1) quasar name; (2) ELG candidate ID; (3) peak
            wavelength; (4) redshift estimate (assuming emission is
            [O\,{\sc ii}]); (5) TTF continuum magnitude, $I(AB)$ - these are
            denoted as limits equal to the limiting magnitude where
            there is no continuum magnitude; (6) $I$-band magnitudes
            from Barr \etal 2003 or Barr 2003\nocite{barr03,barr03t}
            -- these are left blank where no data are available and
            denoted as limits where there is coverage but no
            detection; (7) line flux; (8) observed equivalent width;
            (9) line luminosity (assuming emission is [O\,{\sc ii}])}

\end{deluxetable*}

\subsection{Completeness}
\label{sec:elginc}

There are two mechanisms which affect the completeness of the ELG
candidate catalogues. The first is that objects with peak intensities
below the magnitude completeness limit remain undetected. This is what
is understood by completeness of catalogues in traditional broad-band
imaging. There is an additional phenomenon in the present work,
associated with the ELG line-fitting selection algorithm. If the line
is not sufficiently distinguished from the continuum, as can happen if
the line is faint, or the continuum bright, the candidate may not be
detected. In this section we estimate the number of candidates missed
by these two processes separately.

\begin{figure*}

  \epsscale{1}
  \plottwo{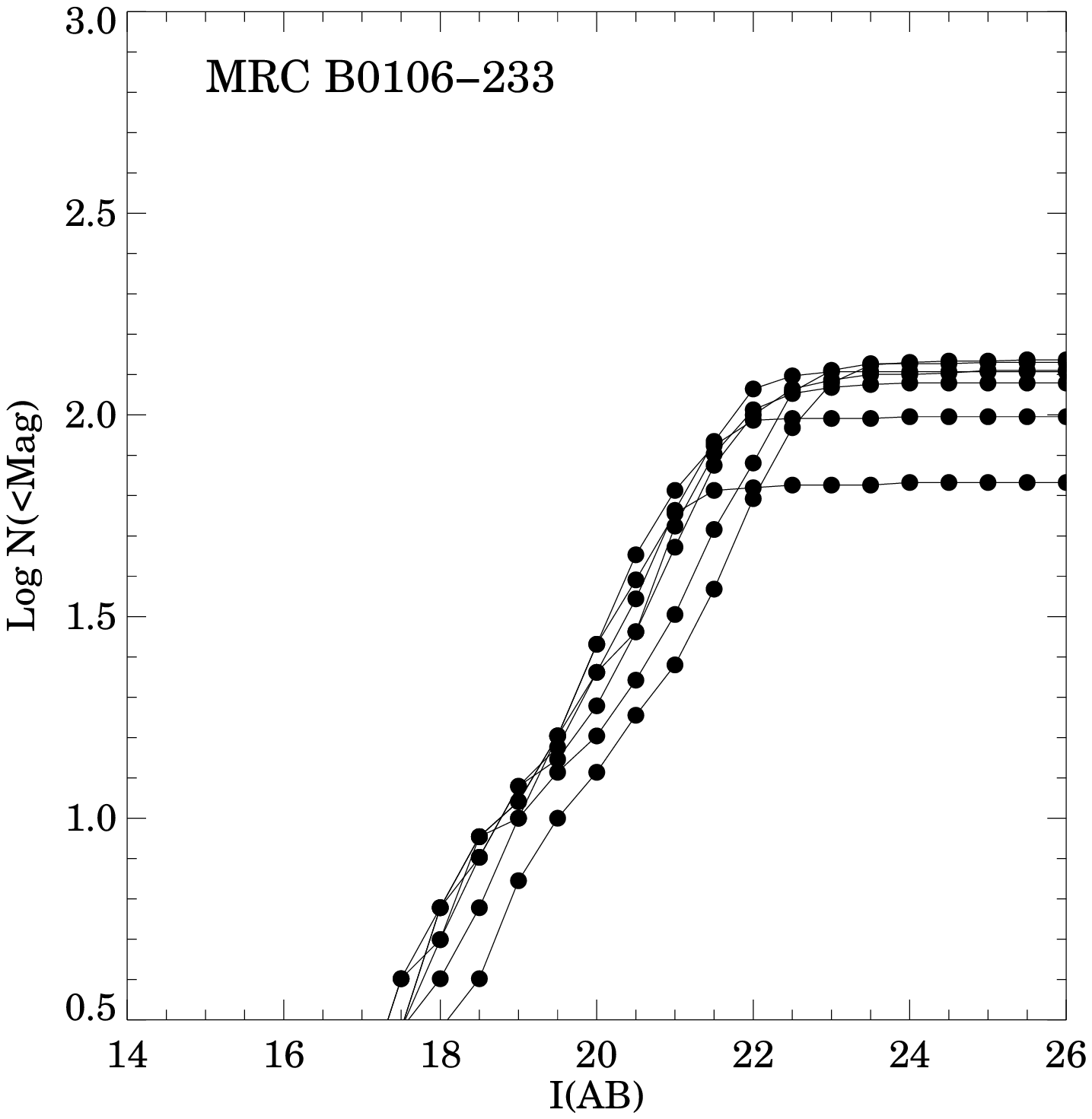}{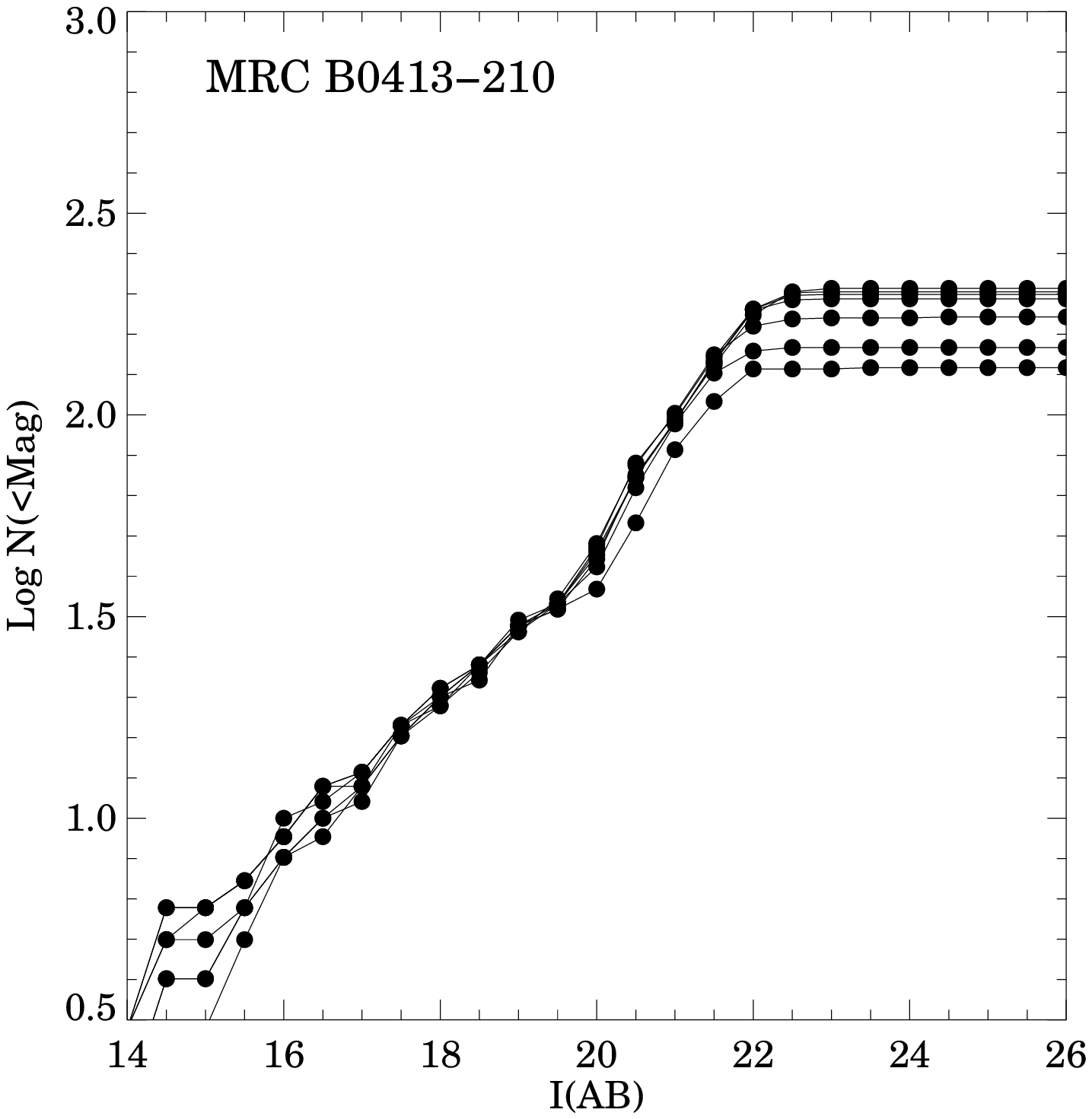}
  \plottwo{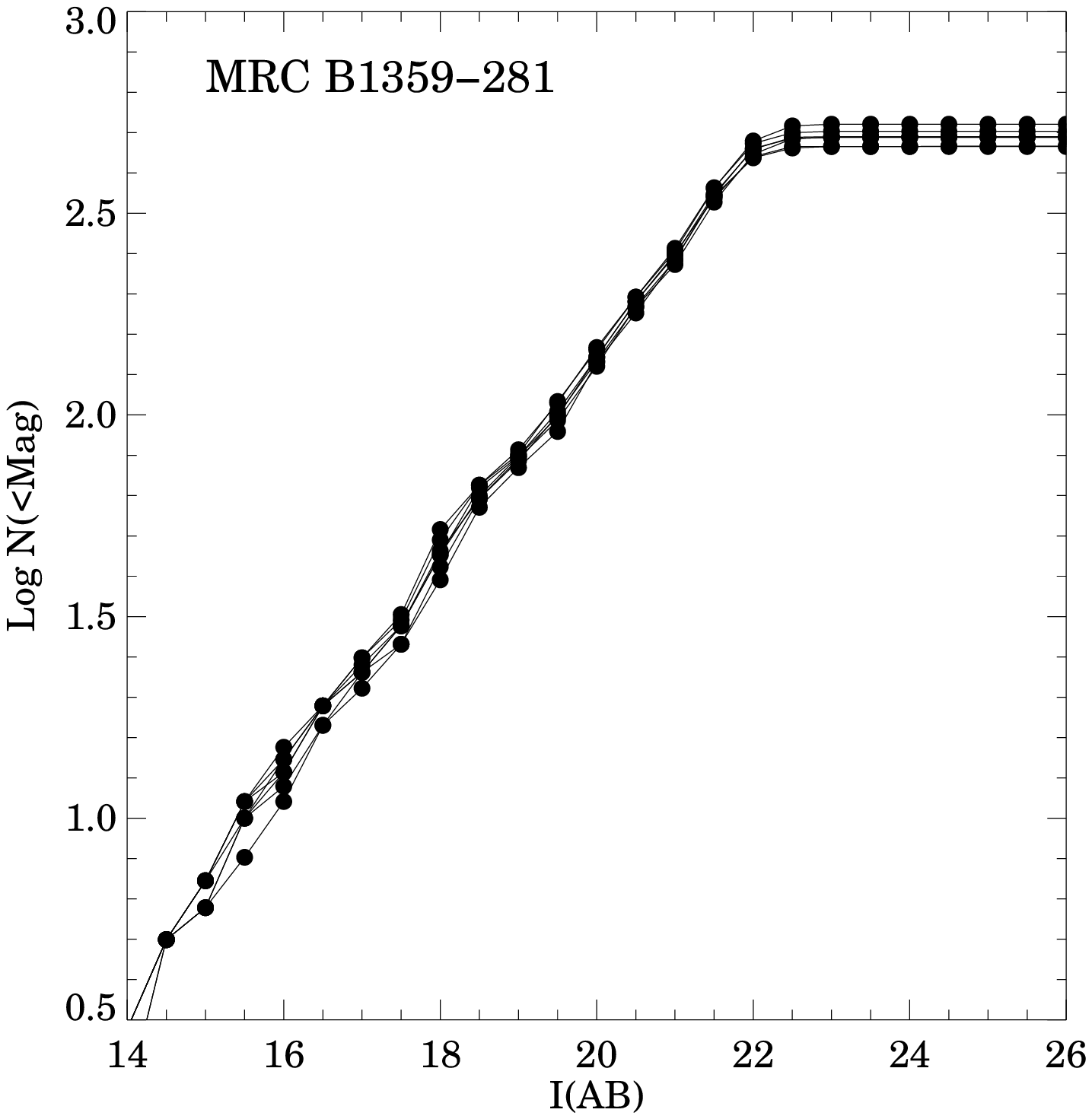}{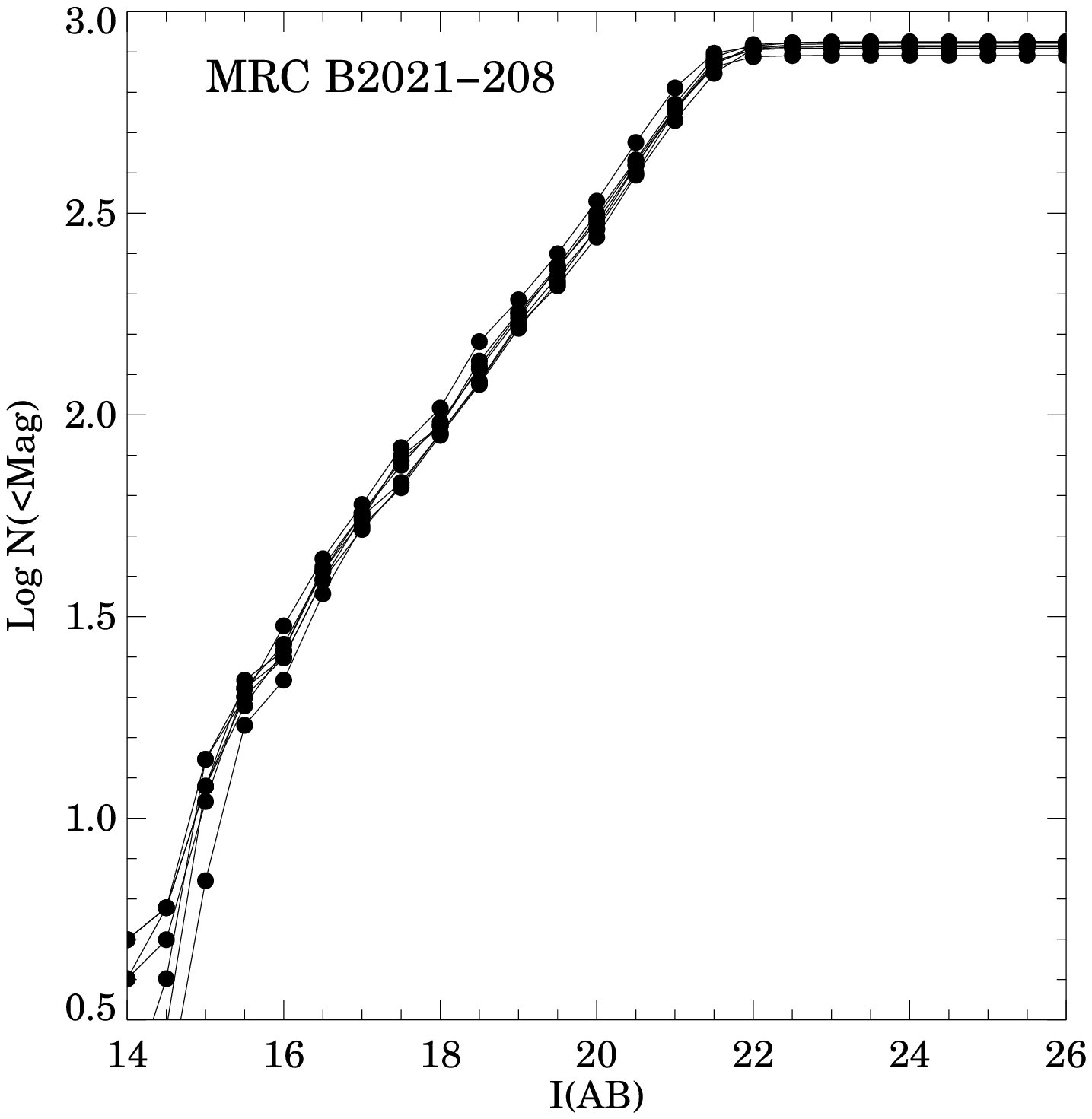}
  \plottwo{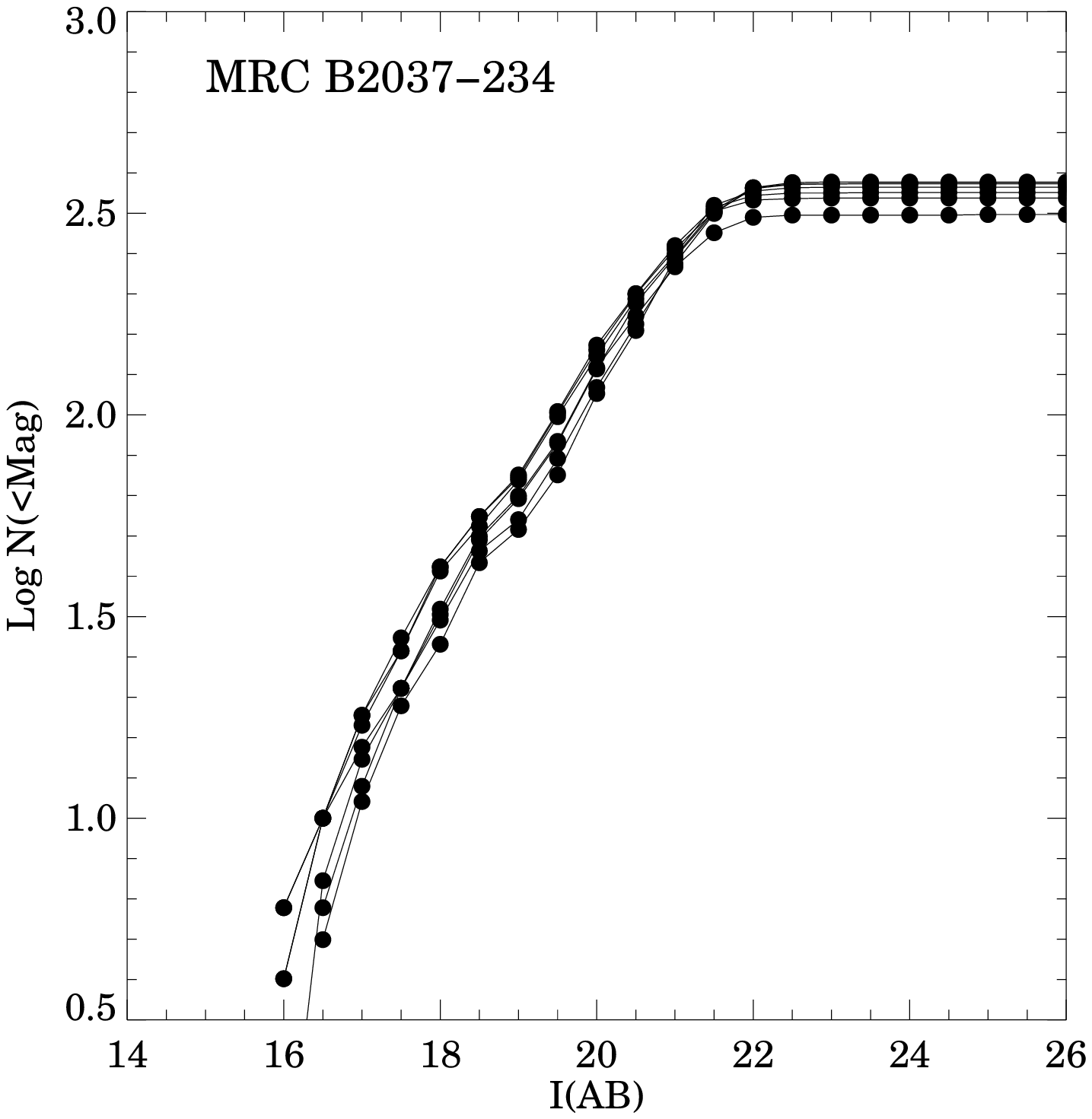}{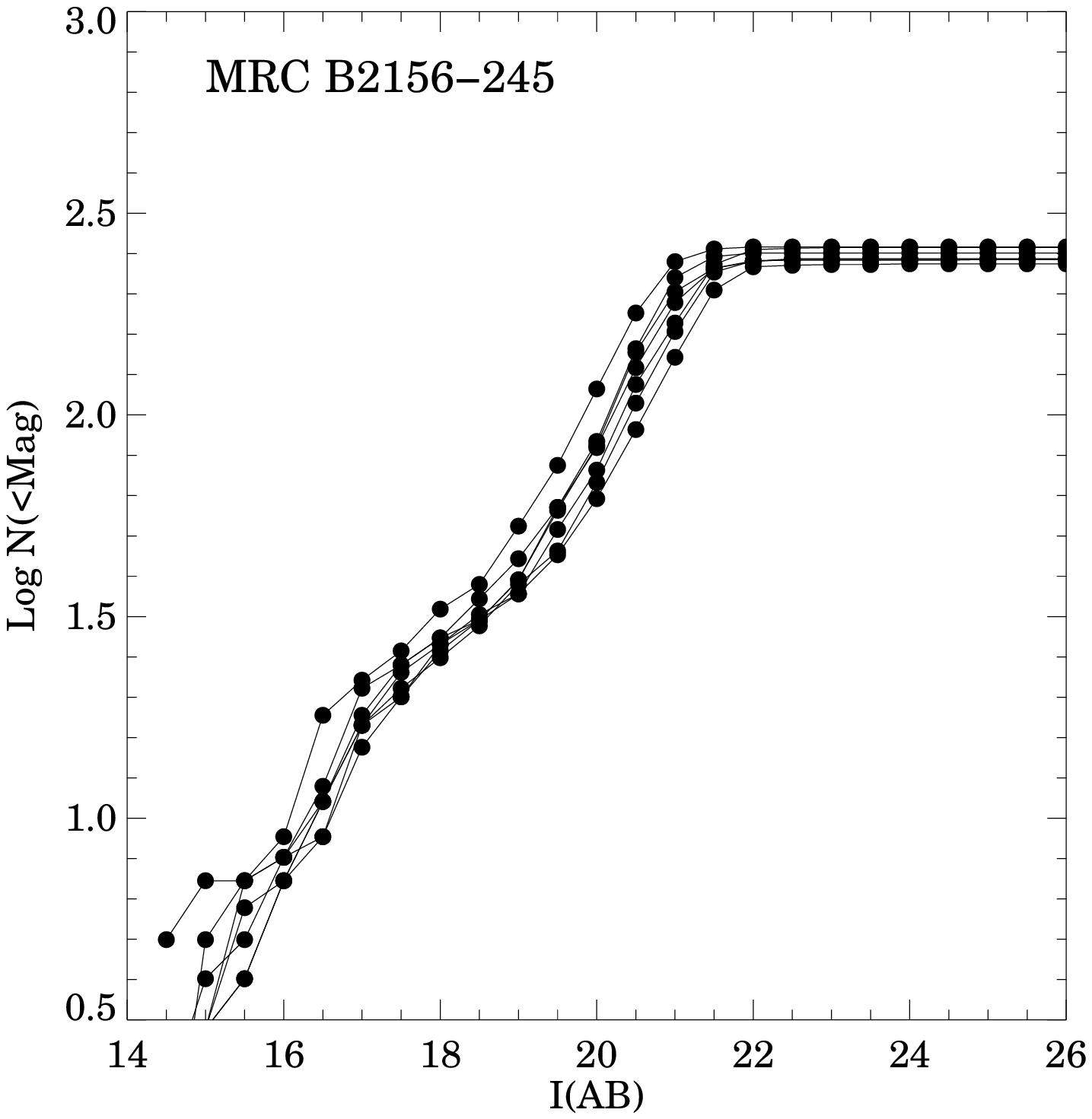}

  \figcaption[Barr.fig5a.eps,Barr.fig5b.eps,Barr.fig5c.eps,Barr.fig5d.eps,Barr.fig5e.eps,Barr.fig5f.eps]{Number
  of objects extracted at each etalon value in each field brighter
  than a given $I(AB)$ magnitude.\label{fig:com}}
  
\end{figure*}

The cumulative number of all objects detected per half magnitude at
each of the seven etalon values in each field are shown in
Figure~\ref{fig:com}. These show that, with the exception of
MRC~B0106--233 which has substantially lower throughput in its reddest
bands (see Figure~\ref{fig:q0413} and \S\ref{sec:0106}), the
completeness in a particular field are the same for all etalon
values.

By extrapolating from the panels in Figure~\ref{fig:com} and assuming
a constant ratio between ELG candidates and other objects, per
luminosity bin, we can estimate how many line emitters remain
undetected because they fall below the completeness limit of the
data. This analysis indicates that 7 candidates are missed in the
field of MRC~B0413--210 and 3 in the field of MRC~B2037--234. The
catalogues in all other fields are complete in that none of the
candidates appear below the completeness limit of the data.

In order to test the algorithms which select objects
based on fluxes above a fitted threshold, artificial catalogues were
created. For each field $10000$ objects were simulated over a range of
fluxes. These line emitters were created by deconvolving a Lorentzian
curve with arbitrary peak wavelength and FWHM equal to the effective
bandpass into seven bins. The resulting spectra were combined with the
observational setup of each scan. The sky noise, efficiency and
wavelength sensitivity were folded in to give catalogues of ELG
candidates to submit to the selection algorithms.

The results are shown in Table~\ref{tab:int} indicating that the
selection algorithm is fairly robust. One line emitter is detected
below the $50\%$ completeness level (in the field of MRC~B0413--210),
which suggests that one object is missed due to failings in the
selection algorithm in this field. A similar analysis for the other
five fields shows that no other candidates remain undetected.

\begin{deluxetable*}{c|rcc|c|cccccc}

  \tablewidth{0pt} \tablecaption{Selection-algorithm completeness
  and predicted rates of interloper detections for different line
  emitters in the fields of the RLQs.\label{tab:int}}
  \tabletypesize{\scriptsize} \tablecolumns{11}

  \tablehead{\colhead{MRC} & \multicolumn{3}{|c|}{Selection-algorithm
   completeness} & \colhead{} & \multicolumn{4}{c}{Interloping line
   emission} & \colhead{} & \colhead{$N_{ELG}$} \\
   \multicolumn{1}{c|}{quasar} & \colhead{$\;\, 100\%$} &
   \colhead{$75\%$} & \multicolumn{1}{c|}{$50\%$} & \colhead{} &
   \colhead{[O\,{\sc ii}]} & \colhead{H$\alpha$} &
   \multicolumn{2}{c}{[O\,{\sc iii}]} & \colhead{} & \colhead{} \\
   \colhead{} & \multicolumn{3}{|c|}{($\times 10^{-16}$ erg s$^{-1}$
   cm$^{-2}$)} & \colhead{} & \colhead{$\lambda 3727$} &
   \colhead{$\lambda 6563$} & \colhead{$\lambda 5007$} &
   \colhead{$\lambda 4959$} & \colhead{} & \colhead{}
  }

  \startdata

     & & & & $z$ & 0.818 & 0.032 & 0.353 & 0.366 \\ B0106--233 & 1.2
     (4) & 0.3 (1) & 0.2 (0) & $V$ & 411 & 3 & 177 & 186 \\ & & & &
     $N_{s}$ & 0.6 & $ < 0.1$ & $<1$ & $<1$ & & 5 \\

   \tableline

    & & & & $z$ & 0.807 & 0.026 & 0.345 & 0.358 \\ B0413--210 & 0.4
    (2) & 0.3 (1) & 0.3 (1) & $V$ & 551 & 3 & 233 & 244 \\ & & & &
    $N_{s}$ & 0.9 & $ < 0.1$ & $<1$ & $<1$ & & 11 \\

   \tableline

    & & & & $z$ & 0.802 & 0.023 & 0.341 & 0.354 \\ B1359--281 & 0.1
     (0) & 0.1 (0) & 0.1 (0) & $V$ & 655 & 3 & 274 & 287 \\ & & & &
     $N_{s}$ & 1.0 & $<0.1$ & $<1$ & $<1$ & & 10 \\

   \tableline

    & & & & $z$ & 1.299 & 0.306 & 0.711 & 0.728 \\ B2021--208 & 0.3
     (1) & 0.2 (0) & 0.2 (0) & $V$ & 709 & 206 & 522 & 532 \\ & & & &
     $N_{s}$ & 0.5 & 0.7 & $<2$ & $<2$ & & 7 \\

   \tableline

    & & & & $z$ & 1.15 & 0.22 & 0.60 & 0.62 \\ B2037--234 & 0.5 (1) &
    0.4(0) & 0.3 (0) & $V$ & 759 & 141 & 500 & 515 \\ & & & & $N_{s}$ &
    0.5 & 0.4 & $<2$ & $<2$ & & 5 \\
 
   \tableline

    & & & & $z$ & 0.862 & 0.057 & 0.386 & 0.399 \\ B2156--245 & 0.5
     (1) & 0.4 (0) & 0.3 (0) & $V$ & 637 & 13 & 298 & 310 \\ & & & &
     $N_{s}$ & 1.0 & $ < 0.1$ & $<2$ & $<2$ & & 9

  \enddata

  \tablecomments{Selection-algorithm completeness indicates the
  efficiency of the ELG selection. The columns describe fluxes, in
  units of $10^{-16}$ erg s$^{-1}$ cm$^{-2}$, where the catalogues of
  artificial objects are $100\%$, $75\%$ and $50\%$ complete. Numbers
  of ELG candidates detected by TTF below this limit are indicated in
  parentheses. $z$: Redshift of line; $V$: Total volume surveyed at
  this redshift ($h^{3}_{70}$ Mpc$^{3}$); $N_{s}$: Expected number of
  interlopers; $N_{ELG}$: Number of ELG candidates found.}

\end{deluxetable*}

There are several reasons why the completeness is different for each
field. The photometric conditions and seeing in each scan is
different; the sky brightness varies according to the wavelength; the
increasing prevalence of OH emission lines at longer wavelengths makes
the sky noisier and ELG detection consequently becomes more
difficult. It is estimated that 10 ELG candidates are missed because
their flux is below the completeness limit of the
catalogue. Simulations of the ELG detection algorithm indicate that a
single candidate is missed by this selection. Caution must be
exercised before we add candidates to our sample, or enhance our
statistics. The estimate of the number of objects missed depends on
the presumption that the luminosity function of line emission traces
that of the underlying population (as sampled by
Figure~\ref{fig:com}). Because of the uncertainty inherent in this
estimate, and because the actual brightness distribution of ELG
candidates is not clear, we take a conservative approach and make no
corrections for incompleteness in the analysis that follows.

\subsection{Possible contaminants}
\label{sec:elgspu}

Within the wavelength range of the observations, line emission from
lower-redshift, star-forming galaxies may be observed. The strongest
lines observed in these galaxies are generally due to
H$\alpha$\,$\lambda 6563$, [O\,{\sc ii}]\,$\lambda 3727$ and [O\,{\sc
iii}]\,$\lambda\lambda 4959,5007$ and an estimate of the number of
these interlopers is required before we evaluate our statistics.

We use the censuses of Gallego \etal (1995; $z<0.1$)\nocite{gallego95}
and Cowie \etal (1997; $z > 0.1$)\nocite{cowie97} to provide estimates
of the `field' density of star-forming galaxies at a given epoch. The
numbers per unit SFR are converted to numbers per unit line luminosity
using the empirical relationships for H$\alpha$
\citep{kennicutt92,kennicutt98} and [O\,{\sc ii}]
\citep{gallagher89,kennicutt98}. The projected number of [O\,{\sc
iii}] emitters is harder to estimate because [O\,{\sc iii}] strength
is only poorly correlated with SFR. However, assuming that [O\,{\sc
iii}] is similar in luminosity to [O\,{\sc ii}] in star-forming
galaxies, we derived estimated numbers. These should be thought of as
upper limits as in most ELGs [O\,{\sc iii}] is weak in comparison to
[O\,{\sc ii}]. Corresponding analyses can be undertaken for other
emission lines (\eg H$\beta$, H$\delta$, etc). However, lines other
than H$\alpha$, [O\,{\sc ii}] and [O\,{\sc iii}] are faint, so only
the strongest star-formers will contribute. These are rare and the
fraction of ELGs of this type found serendipitously is likely to be
negligible. The numbers of interlopers predicted by this method for
each field are shown in Table~\ref{tab:int}.

The numbers given in Table~\ref{tab:int} for [O\,{\sc ii}] and H$\alpha$
emitters will be uncertain by a factor of a few. This is because of
the scatter in the empirical luminosity-SFR relationship and the large
uncertainties (at least a factor of 2) in the data of Cowie \etal The
number of interlopers shown in Table~\ref{tab:int} is further
overestimated because of the restricted magnitude range of the
detections ($21 < I(AB) \lesssim 23$). 

At high redshifts, the deep survey of Rhoads \etal
(2000)\nocite{rhoads00} predicts $\sim 4000$ deg$^{-2}$ z$^{-1}$
Ly$\alpha$ emitters in blank field surveys. This amounts to $\sim 1$
per TTF field. However, this figure is based on the detection of a
single galaxy at $z=4.52$ with line flux of $1.7 \times 10^{-17}$ erg
s$^{-1}$ cm$^{-2}$ -- below the completeness limits of our
fields. This flux level, combined with the uncertainties inherent in
extrapolating from a single object, and the high cosmic variance of
Ly$\alpha$ emitters, makes it very difficult to estimate the number of
such objects we should see. We therefore make no correction for
contaminant Ly$\alpha$ emitters.

As well as star-forming galaxies, our search may detect line emission
from AGN. However, AGN have a much lower co-moving number density than
emission-line galaxies (Grazian \etal 2000; Hicks \etal
2002)\nocite{grazian00,hicks02,wang04}. This means that their
contamination rate will not significantly affect the predictions for
interlopers made above.

These estimates assume that we can treat star-forming galaxies as a
homogeneously distributed population, which, of course, is not the
case. ELGs are often found as multiples (\eg Hutchings \etal 1993;
Hicks \etal 2002; Venemans \etal
2002)\nocite{hutchings93,hicks02,venemans02}. However, accounting for
this variance is very difficult as the clustering of star-formers over
a range of redshift is not clearly mapped. The predicted rate of
interloping H$\alpha$ for most fields is $<0.1$, the expected number
of [O\,{\sc iii}] emitters is an upper limit based on the assumption
that [O\,{\sc iii}] traces SFR in the same way as [O\,{\sc ii}], and
imposing a magnitude cut at $I(AB) > 21$ will cull lower redshift line
emitters. Therefore, we expect objects detected by our TTF
observations primarily to be [O\,{\sc ii}] emitters.

The wavelength regions probed in this paper do not contain any 
sharp features that might arise in Galactic stars, so stars
will not be a significant contaminant. We note that the wavelength
range targetted in Paper 1 was clipped as a precaution against 
possible contamination by M-stars.

\subsection{Notes on individual fields}
\label{sec:elg_dis}

\subsubsection{MRC~B0106--233}

\label{sec:0106}

No evidence is seen for [O\,{\sc ii}] emission from the quasar, consistent
with the published nuclear spectrum \citep{baker99}. The analysis of
this field is inhibited by the response of the blocking filter which
drops off substantially in the red. This is less of a problem than
appears to be the case from Figure~\ref{fig:ban} because only the
central parts of each image sample the reddest wavelengths. An
examination of the number of objects detected by SExtractor indicates
that noticeably fewer objects are found at the reddest three $Z$
values. These are removed from any further analysis.

\begin{figure}

  \epsscale{1}\plotone{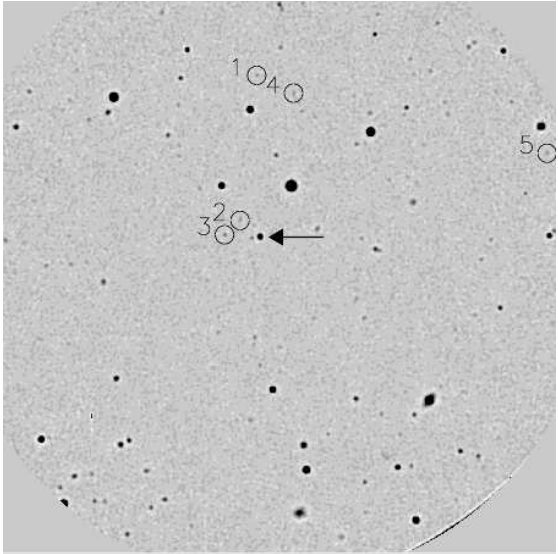}

  \figcaption[Barr.fig6.eps]{The distribution of ELG candidates in the
  field of MRC~B0106--233. The field of view is \arcmn{7}{4} $\times$
  \arcmn{7}{4} of a composite of the seven (circular) TTF images.  The
  quasar is marked with the horizontal arrow; North is up, East is
  left.\label{fig:0106_loc}}

\end{figure}

Five ELG candidates are picked out from what is, in terms of total
objects, a sparse field (Figure~\ref{fig:0106_loc}). Two of the best
candidates in the field (IDs 2 and 3) are situated within
$10^{\prime\prime}$ of the quasar. Their emission also coincides with
the quasar redshift, both having $z = 0.817 \pm 0.002$. Note that
Figure~\ref{fig:0106_loc} and the corresponding figures for other
fields are created by combining the seven narrow-band images. They can
therefore be considered as images taken through a passband of 70 --
100\AA \ covering the redshifted [O\,{\sc ii}]. ELG candidates in these images
appear brighter than the continuum magnitude documented in columns 5
and 6 of Table~\ref{tab:elg_1} because the figures include, and indeed
isolate, the line emission.

\subsubsection{MRC~B0413--210}

\label{sec:0413}

The quasar brightens blueward of the position expected for [O\,{\sc
ii}] at $z=0.807$. However, the TTF wavelength calibration was taken
nine hours before the data scan during which time the $Z,\lambda$
relation may have drifted. The redshift is originally determined from
C\,{\sc iii}], C\,{\sc iii}] and Mg\,{\sc ii} and does not catalogue
the observed wavelength of [O\,{\sc ii}] \citep{wilkes86}. For the
purposes of this work it is assumed that the quasar is at $z = 0.807$
and the peak flux in the TTF image of the RLQ is due to [O\,{\sc ii}]
emission. The wavelength solution was therefore adjusted 12\AA \
redward to correct for the discrepancy. Note that this recalibration
affects the wavelength space, not the amount of volume sampled or the
flux calibration. The only analysis thus affected is the distribution
of ELGs in velocity space. Figure~\ref{fig:elg_v} illustrates the
effect that the adjustment has on the space density \vs velocity
histogram.

\begin{figure*}

  \epsscale{1}\plotone{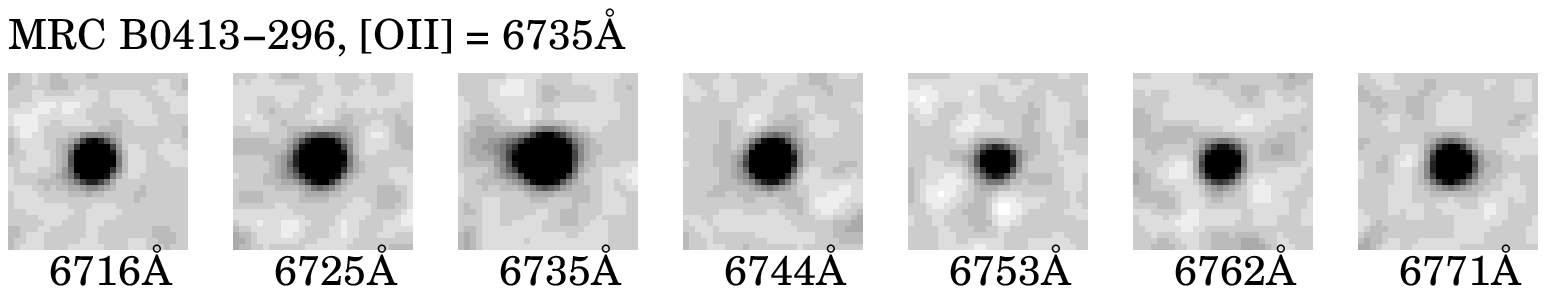}

  \figcaption[Barr.fig7.eps]{TTF images of MRC~B0413--210 in each
  band. The field of view in each panel is \arcsd{11}{1} $\times$
  \arcsd{11}{1} and the central wavelength is indicated at each
  panel.\label{fig:q0413}}

\end{figure*}

Assuming that the peak in the quasar brightness is due to [O\,{\sc
ii}], the flux response is that indicated in
Figure~\ref{fig:q0413}. The morphology of the line emission is
elongated slightly in the E -- W direction, more-or-less aligned with
the radio emission \citep{kapahi98}. However, the images in
Figure~\ref{fig:q0413} have mediocre spatial resolution (seeing
$\approx$ \arcsd{1}{0}). Nor is it out of the question that the
`alignment' is caused by a faint ELG located just East of the
quasar. MRC~B0413--210 is a small radio source (largest angular size
$\sim 5^{\prime\prime}$) and the highest resolution radio map in
Kapahi \etal (1998)\nocite{kapahi98} is not detailed. Higher
resolution radio and optical narrow-band imaging will be needed to
ascertain the veracity of this alignment effect.

\begin{figure}

  \epsscale{1}\plotone{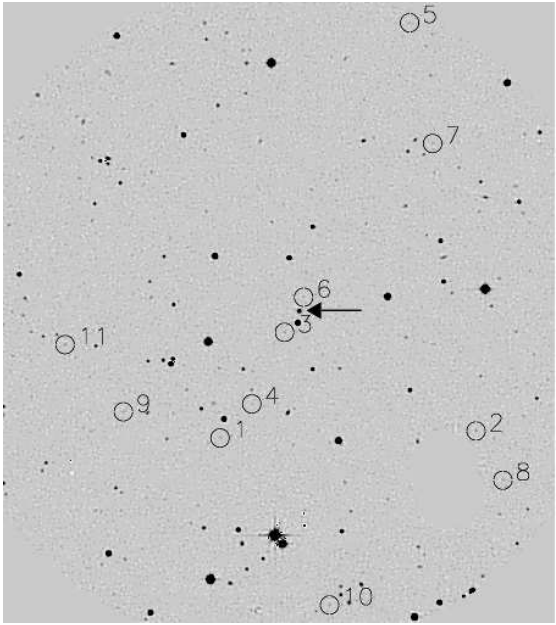}

  \figcaption[Barr.fig8.eps]{The distribution of ELG candidates in the
  field of MRC~B0413--210. The image is a composite of the seven TTF
  bands cut to \arcmn{7}{4} $\times$ \arcmn{8}{3} about the quasar
  (marked with the horizontal arrow). North is up, East is
  left.\label{fig:0413_loc}}

\end{figure}

None of the eleven ELG candidates detected in this field is
particularly strong, with only one object having an intrinsic
equivalent width $>50$\AA, and just one other with a detectable
continuum magnitude (object 11). As with the field of MRC~B0106--233
there is the coincidence of a pair of candidates with the quasar's
spatial position. There is also overlap in their redshifts, which are
$z = 0.805 \pm 0.002$ and $z = 0.808 \pm 0.002$, well matched to the
quasar redshift. There is no further obvious clustering of ELG
candidates in the field of MRC~B0413--210 (Figure~\ref{fig:0413_loc}).

\subsubsection{MRC~B1359--281}

The quasar shows a strong peak in brightness at the wavelength of
[O\,{\sc ii}] at $z=0.802$, although its morphology remains optically
unresolved. In radio terms, MRC~B1359--281 is a compact steep-spectrum
source (see Kapahi et al. 1998) \nocite{kapahi98}.

\begin{figure}

  \epsscale{1}\plotone{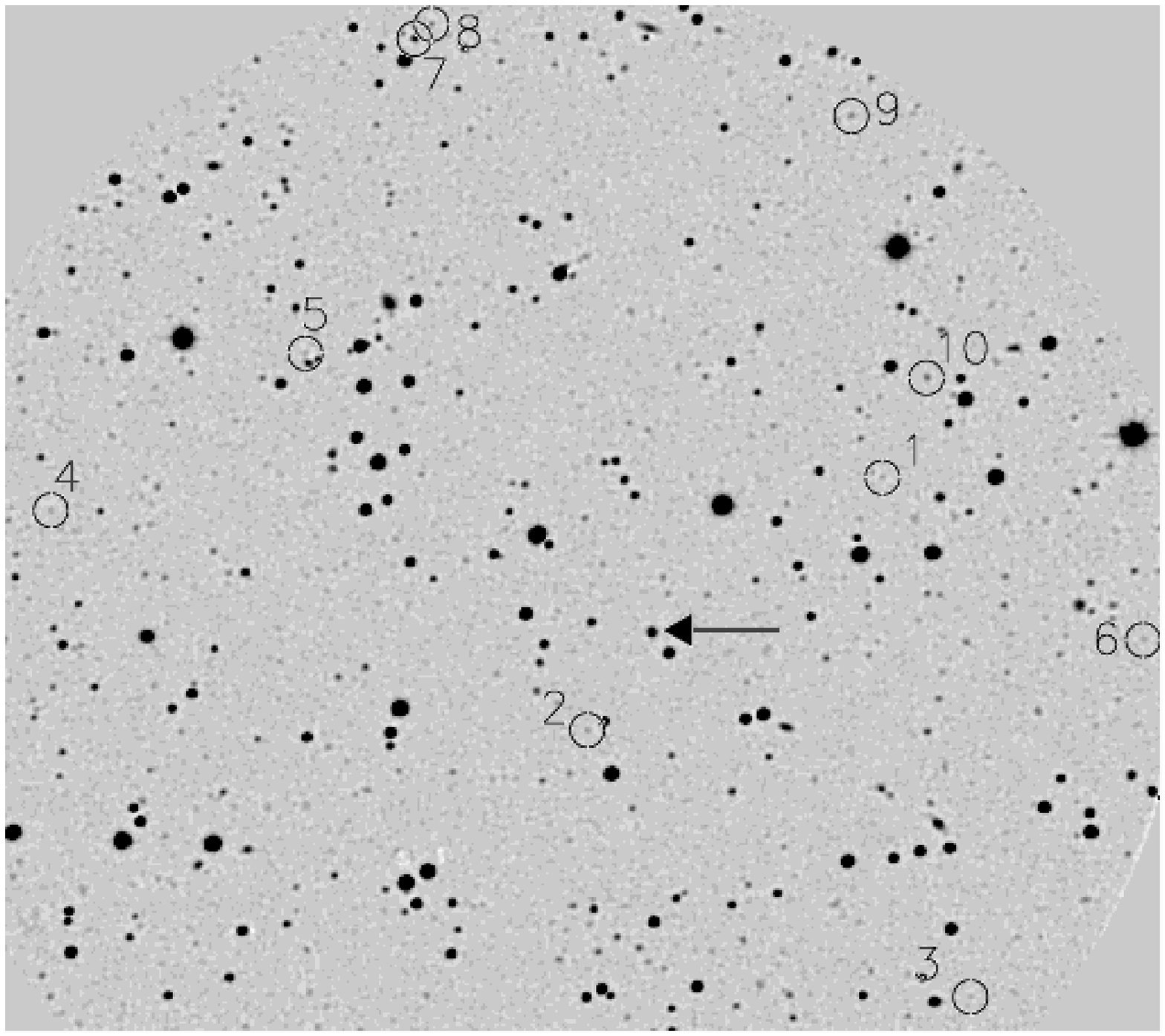}

  \figcaption[Barr.fig9.eps]{The distribution of ELG candidates in the
  field of MRC~B1359--281. The field of view is \arcmn{8}{3} $\times$
  \arcmn{7}{4} of a composite of the seven TTF bands. The quasar is
  marked with the arrow. North is up, East is
  left.\label{fig:1359_loc}}

\end{figure}

Ten ELG candidates are detected, including seven objects with
$W_{\lambda} > 50$\AA. Apart from a pair $\sim 5^{\prime}$ from the
quasar, there is no clustering of these objects
(Figure~\ref{fig:1359_loc}).

The quasar was imaged as part of a broad-band program to detect
clustering of passively-evolving ellipticals in the fields of RLQs
(Barr \etal 2003)\nocite{barr03}. No evidence for a group or cluster
of red galaxies was found.

\subsubsection{MRC~B2021--208}
\label{sec:2021}

There is no significant peak in the quasar flux at 8568\AA, the
wavelength expected of [O\,{\sc ii}] at $z=1.299$. Nor does the
morphology change from that of an unresolved point source. No obvious
[O\,{\sc ii}] emission is seen in the published spectrum (Murdoch,
Hunstead \& White 1984)\nocite{murdoch84}.

Two objects with continuum magnitude brighter than the $I(AB) = 21$
magnitude cutoff are found to display very strong line emission. A
first-ranked cluster member at the redshift of the quasar would have
$I \sim 21 - 22$ \citep{eales85,snellen96}. Emission-line galaxies at
$z = 1.299$ should be fainter than this, so it seems likely that ELG
candidates 5 and 6 are not [O\,{\sc ii}] emitters at $z = 1.299$. The
strongest expected line associated with ELGs is H$\alpha$ which the
TTF setup in the present case is sensitive to at $z = 0.30$. The
number of H$\alpha$ interlopers within the volume of observation is
predicted to be $\sim 1$. It is therefore likely that the line
emitters with $I < 21$ are H$\alpha$ sources at $z = 0.3$.

\begin{figure}

  \epsscale{1}\plotone{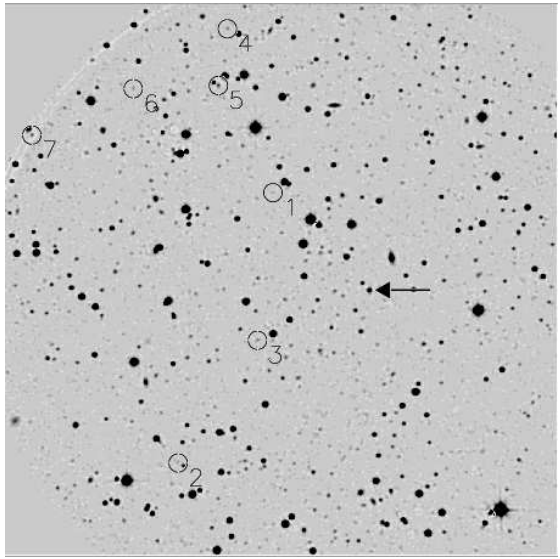}

  \figcaption[Barr.fig10.eps]{The distribution of ELG candidates in
  the field of MRC~B2021--208. The field of view is \arcmn{7}{4}
  $\times$ \arcmn{7}{4}. The image is a composite of the seven TTF
  bands and the quasar is marked with a horizontal arrow. North is up,
  East is left.\label{fig:2021_loc}}

\end{figure}

In terms of spatial position, the ELG candidates locate themselves
primarily to the NE of the quasar
(Figure~\ref{fig:2021_loc}). However, the spatial clustering is rather
loose and three of these objects are the brightest in the field. If
these are indeed not [O\,{\sc ii}] emitters, they may be a group of H$\alpha$
emitters at $z = 0.30$. Deep spectroscopy will be required to test
this hypothesis. 

\subsubsection{MRC~B2037--234}

The quasar is faint, with magnitude $I(AB) \approx 21$. This makes it
difficult to characterise any [O\,{\sc ii}] emission, nuclear or extended.
No strong [O\,{\sc ii}] emission from MRC~B2037--234 is seen. 
There is no published optical spectrum for this source and the
redshift is more uncertain than the other quasars observed in this
paper.

\begin{figure}

  \epsscale{1}\plotone{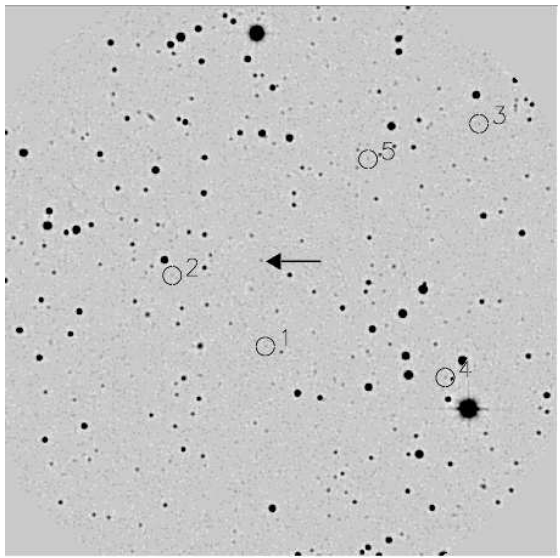}

  \figcaption[Barr.fig11.eps]{The distribution of ELG candidates in
  the field of MRC~B2037--234. The field of view is \arcmn{7}{4}
  $\times$ \arcmn{7}{4} of a composite of the seven TTF bands and the
  quasar is marked by an arrow. North is up, East is
  left.\label{fig:sp37}}

\end{figure}

Only five ELG candidates are detected in this field. None is an
especially strong line emitter, nor is any one found at the wavelength
expected for [O\,{\sc ii}] at $z = 1.15$.  As indicated in
Table~\ref{tab:int}, there is no clear excess of ELG candidates in
this field at the quasar redshift.  The numbers are consistent with
the expected number of interlopers and field galaxies. The candidates
are spread across the field, with no sign of clustering
(Figure~\ref{fig:sp37}).

\subsubsection{MRC~B2156--245}

The quasar brightness peaks at 6949\AA, $\sim 9$\AA \ redward of the
wavelength of [O\,{\sc ii}] at $z=0.862$, which may indicate that the
$Z,\lambda$ relation has drifted. Indeed, the wavelength calibration
used in this field comes from the previous night. However, the
published spectrum of MRC~B2156--245 \cite{baker99} has [O\,{\sc ii}]
at 6952\AA, consistent with the TTF result. No adjustment to the
wavelength solution is made.

\begin{figure*}

  \epsscale{1}\plotone{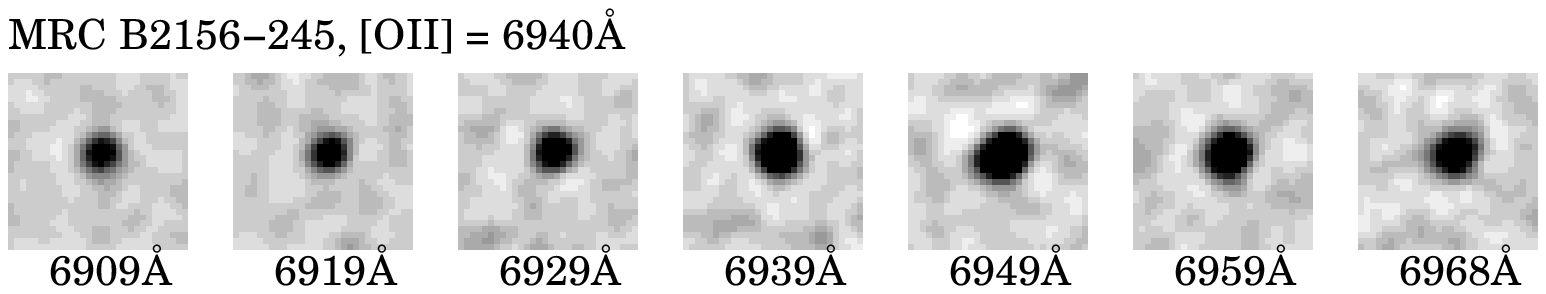}

  \figcaption[Barr.fig12.eps]{TTF images of MRC~B2156--245 in each
  band. The field of view in each panel is \arcsd{11}{1} $\times$
  \arcsd{11}{1} and the central wavelength is indicated at each
  panel. \label{fig:q2156}}

\end{figure*}

The [O\,{\sc ii}] emission from MRC~B2156--245 appears to align itself
$\sim$ NW -- SE (Figure~\ref{fig:q2156}). Recent MERLIN images show
sub-arcsecond mini-lobes extending NW and E of the quasar nucleus (de
Silva et al., in preparation).

\begin{figure}

  \epsscale{1}\plotone{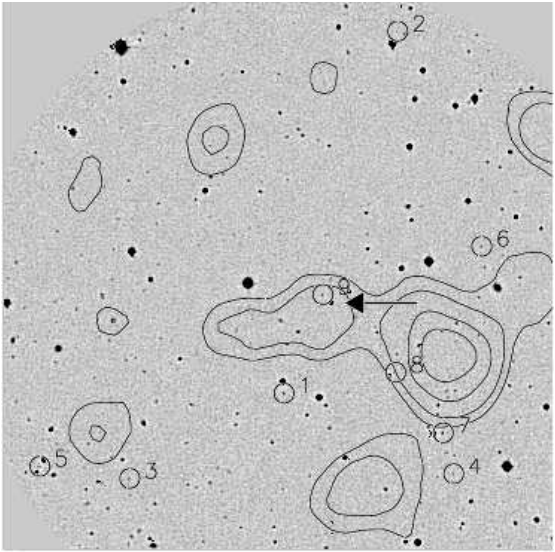}

  \figcaption[Barr.fig13.eps]{The distribution of ELG candidates in
  the field of MRC~B2156--245. The field of view is \arcmn{7}{4}
  $\times$ \arcmn{7}{4} and the quasar is marked with the horizontal
  arrow. The image is a part of composite of the seven TTF bands.
  North is up, East is left. Contours represent the surface density of
  galaxies with colours of passively-evolving elliptical galaxies at
  $z=0.862$ from Barr (2003). The levels are 2, 3, 5 and 7 $\times$
  the Poissonian noise of similarly-coloured
  galaxies.\label{fig:2156_loc}}\nocite{barr03t}

\end{figure}

There are a number of strong line emitters found in this field. Seven
out of nine have emission consistent with [O\,{\sc ii}] at $z
=0.862$. Moreover, they appear to congregate to the SW of the
quasar. This is the site of a cluster of galaxies found in Barr
(2003)\nocite{barr03t}. Figure~\ref{fig:2156_loc} shows that, while
the ELG candidates appear near the red galaxy overdensity, they avoid
the site of peak clustering. This supports the view that the edges,
rather than cores of clusters of galaxies, are predominantly the sites
of star formation activity.

\section{Discussion}
\label{sec:dis}

In this paper we have examined the fields of six quasars and find
forty-seven new ELG candidates. The number of candidates about each
quasar varies from the number expected from field surveys (\eg
MRC~B2037--324) to overdensities of $\sim 10$ times (MRC~B0413--210,
MRC~B1359--281). Previous studies have shown that MRC~B2156--245
resides on the edge of a rich cluster of galaxies \citep{barr03t}. The
other five quasars have either not been examined for evidence
of galaxy clustering (due to observing constraints) or have 
environmental richnesses consistent with the field \citep{barr03t,barr03}.

We now consider the ELG properties and environments of 
all seven MQS quasars targetted with TTF, including
the 17 ELG candidates (3 spectroscopically confirmed) detected in the 
field of MRC~B0450--221 at $z=0.9$ described in Paper 1. 
Clear evidence for a rich cluster of galaxies in the MRC~B0450--221
field was presented in Paper 1.

We note that followup spectroscopy is necessary to determine the
nature of the ELG candidates. For the forthcoming discussion we assume
the line is indeed [O\,{\sc ii}] for all candidates except the two brightest
objects in the field of MRC~B2021--208 (see \S\ref{sec:2021}). In
Paper 1, spectroscopic follow-up found one low-redshift interloper out
of seven targets, which agrees with the number expected from the
analysis described in Section \ref{sec:elgspu}.  Therefore, as the
determination of interlopers in this paper is essentially the same as
that in Paper 1, no great diminution of the statistics presented here
is expected.

\subsection{Extended [O\,{\sc ii}] emission around quasar host galaxies}

It is well known that radio-loud AGN at moderate to high redshifts
exhibit strong alignments between their optical-line and radio
emission (\eg McCarthy 1993; Rush \etal 1997; Axon \etal 2000;
Hutchings, Morris \& Crampton
2001)\nocite{mccarthy93,rush97,axon00,hutchings01}. Indeed, extended
[O\,{\sc ii}] emission is common around 3C quasars at comparable
redshifts to the quasars in our sample (\eg Bremer \etal
1992\nocite{bremer92}). We note that evidence of such a phenomenon in
this work exists for two sources (MRC~B0413--210 and MRC~B2156--245).
However, the resolution of each image precludes detailed optical
mapping of these regions.

\subsection{The distribution of ELG candidates about quasars}

\begin{figure}

  \epsscale{1}\plotone{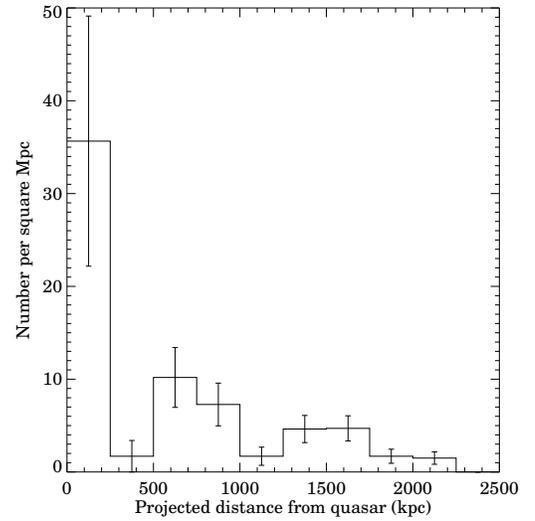}

  \figcaption[Barr.fig14.eps]{The projected distances of ELG
  candidates about the quasars. The $1\sigma$ error bars based on the
  number of objects per bin are shown.\label{fig:elg_dis}}

\end{figure}

Figure~\ref{fig:elg_dis} plots the surface density of candidate [O\,{\sc ii}]
emitters against projected distance from the quasar. There is a peak
in the surface density of ELG candidates within $250$ kpc of the
quasars, and also a smaller overdensity at $500 - 1000$ kpc from the
quasar. The first signature appears to arise from ELG close companions to 
the quasars, e.g. a pair of ELGs lying within $10''$ of MRC~B0106--233, 
and similarly for MRC~B0413--210.
The latter excess occurs at distances consistent with the scale 
lengths of galaxy clusters about AGN at these redshifts 
\cite{hall98,nakata01,bremer02,barr03}.

In the six new fields presented in this paper, no further examples
of strongly clustered ELG groups similar to that seen in the
field of MRC~B0450--221 (Paper 1) are found, apart from the 
quasar companions. Weak clustering of ELG candidates is likely,
but the small numbers preclude a detailed clustering analysis.
For example, ELG candidates are seen solely NE of 
MRC~B2021--208 or SW of MRC~B2156--245.

\begin{figure*}

  \epsscale{1}\plottwo{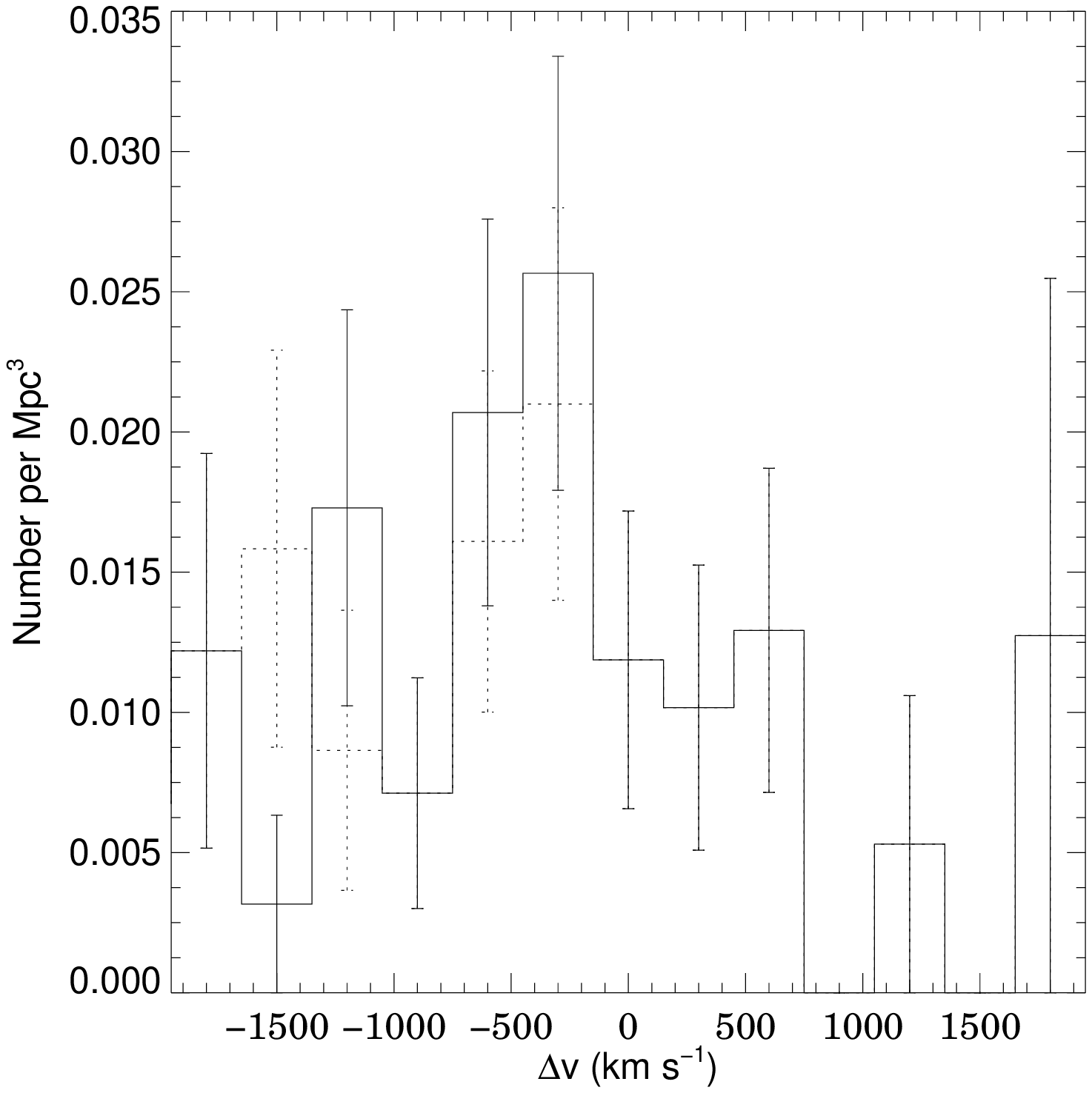}{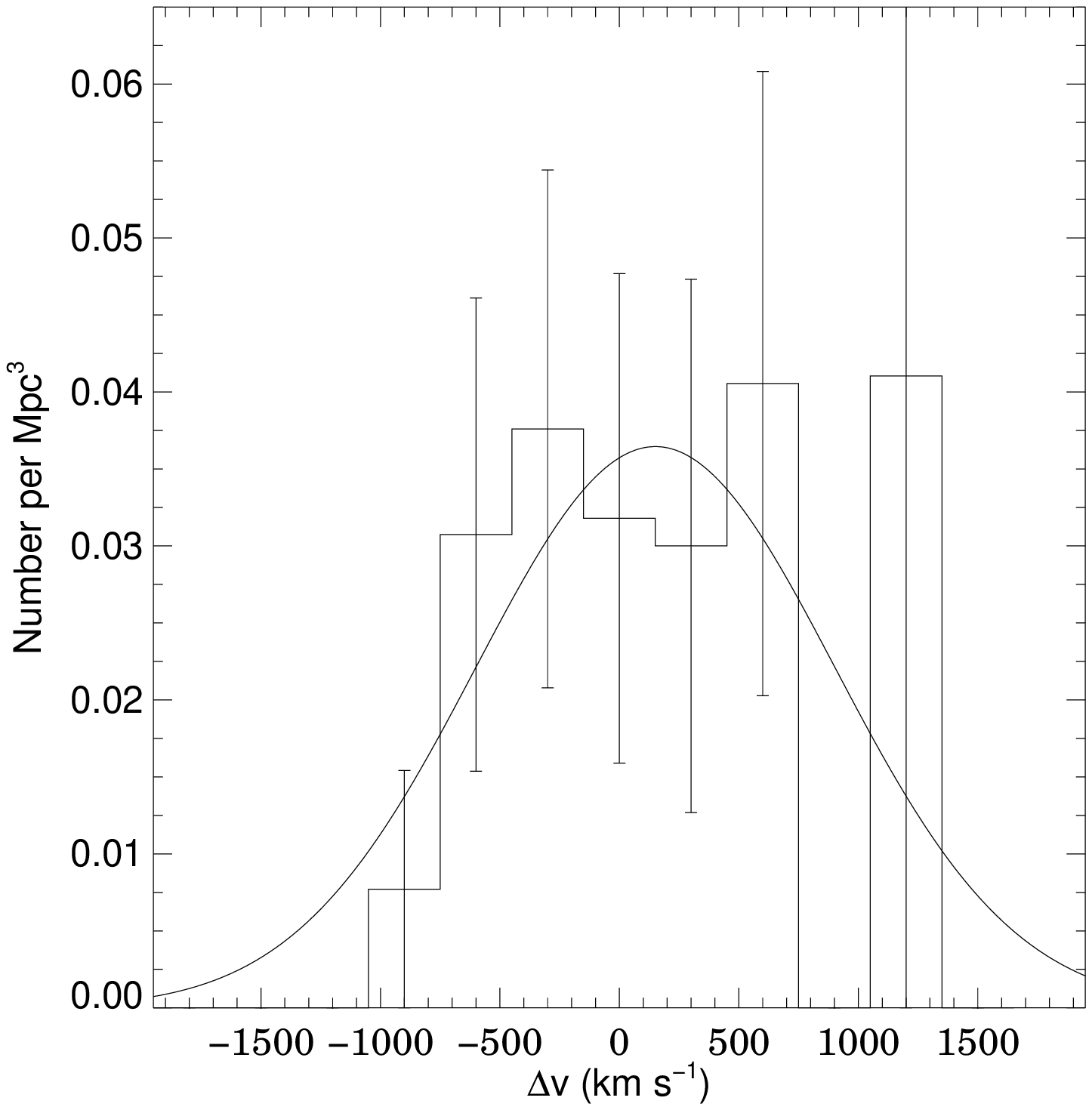}

  \figcaption[Barr.fig15a.eps,Barr.fig15b.eps]{{\em Left:} The space
  density of ELG candidates detected with TTF about quasars in this
  work as a function of velocity relative to the quasar redshift. The
  dashed line shows the distribution before recalibrating the
  wavelength solution from MRC~B0413--210. The expected number of ELGs
  with these properties in the field is expected to be $\lesssim
  0.001$ Mpc$^{-3}$ at these redshifts (Cowie \etal 1997). {\em
  Right:} The same but restricted to two quasars known to reside in or
  near clusters of galaxies (MRC~B0450--221, MRC~B2156--245). The
  solid line denotes the best fit Gaussian curve to the points
  ($\sigma$ is 750 km s$^{-1}$). Error bars are $1\sigma$ based on the
  number of objects per velocity bin; note the different ranges on the
  {\em y}-axes.\label{fig:elg_v}}\nocite{cowie97,baker01}

\end{figure*}

Figure~\ref{fig:elg_v} shows the distribution in velocity space of
[O\,{\sc ii}] candidates detected with TTF about the quasars. The left panel
indicates that the average space density of these objects from $-1000
\lesssim v_{\mathrm{quasar}} \lesssim 500$ km s$^{-1}$ is $\approx
0.01$ Mpc$^{-3}$, ten times greater than the $\lesssim 0.001$
Mpc$^{-3}$ expected from field surveys \citep{cowie97}. The right
panel in Figure~\ref{fig:elg_v} plots the number density \vs velocity
histogram for MRC~B0450--221 and MRC~B2156--245, two quasars known to
reside in rich clusters of galaxies \citep{baker01,barr03t}.  These
fields clearly show a large overdensity of ELG candidates about the
quasar redshift. The spread in velocities seen ($\sigma \approx 750$
km s$^{-1}$, by fitting a simple Gaussian distribution) is comparable
with the average velocity dispersion for Abell richness class 2
clusters of $\sim 800$ km s$^{-1}$ \cite{yee03}.

\subsection{Equivalent widths and star formation rates}

\begin{figure}

  \epsscale{1}\plotone{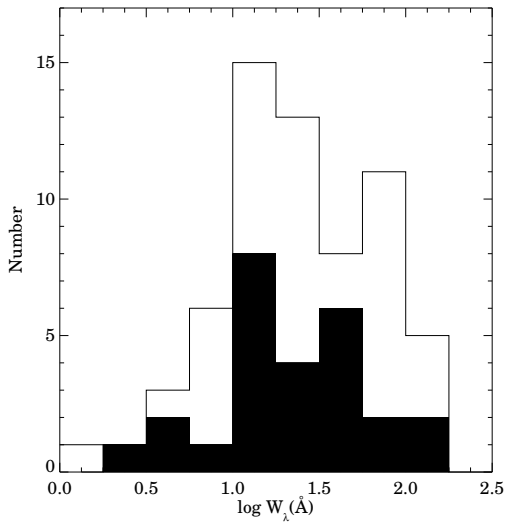}

  \figcaption[Barr.fig16.eps]{Histogram of rest frame equivalent
  widths, $W_{\lambda}$, for ELG candidates in this work.  The
  unfilled parts of the histogram represent those objects without a
  continuum magnitude whose value of $W_{\lambda}$ is a lower
  limit.\label{fig:eqw}}\nocite{baker01}

\end{figure}

The observed equivalent widths for the ELG candidates in this paper
lie in the range $\sim 10 - 200$\AA, corresponding to $\sim 5 -
100$\AA \ in the rest frames of the quasars. Figure~\ref{fig:eqw}
shows the distribution of the intrinsic $W_{\lambda}$. This
distribution of equivalent widths is consistent with the field survey
of Hogg \etal (1998)\nocite{hogg98} of [O\,{\sc ii}] emission-line galaxies in
the field with $0.8<z<1.3$.

\begin{figure}

  \epsscale{1}\plotone{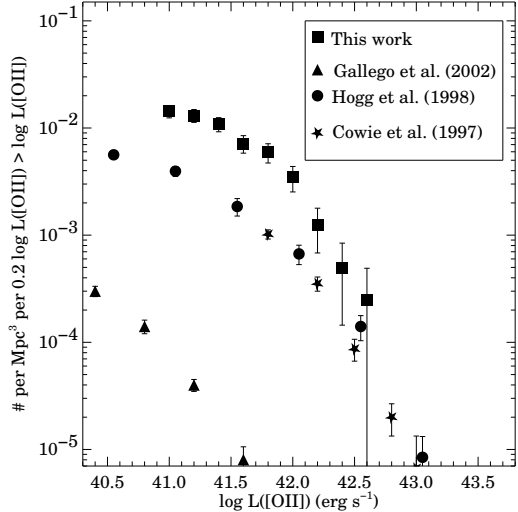}

  \figcaption[Barr.fig17.eps]{Cumulative density of objects brighter
  than a given threshold \vs [O\,{\sc ii}] luminosity. Squares
  represent data points for the quasars in this paper. Circles are
  $0.3<z<1.3$ ELGs from the [O\,{\sc ii}] survey of Hogg \etal (1998).
  Stars represent points with $0.8<z<1.6$ from the Cowie \etal (1997)
  $B$-band survey converted to [O\,{\sc ii}] luminosity using the
  conversion of Gallagher \etal (1989). Triangles represent the local
  [O\,{\sc ii}] luminosity function of Gallego \etal (2002). All
  points have been adjusted to a $H_0 = 70$ km s$^{-1}$ Mpc$^{-1}$,
  $\Omega_{\Lambda}=0.7$ cosmology. Error bars are the root variance
  of the number of objects in each luminosity bin and are therefore
  not
  independent.\label{fig:lumf}}\nocite{hogg98,cowie97,gallagher89,gallego02}

\end{figure}

Figure~\ref{fig:lumf} shows the candidate ELG luminosity function,
assuming that the emission is [O\,{\sc ii}]. Also displayed is the
luminosity function of local [O\,{\sc ii}] emitters \citep{gallego02}
as well as two representations of high-redshift field surveys of
star-forming galaxies \citep{cowie97,hogg98} adjusted to our adopted
cosmology. The number of [O\,{\sc ii}] line emitters per unit
luminosity in the vicinity of quasars at $0.8 \lesssim z \lesssim 1.3$
is $\sim 100$ times greater than those found locally. At $L$([O\,{\sc
ii}]) $< 10^{42}$ erg s$^{-1}$, the density of [O\,{\sc ii}] emitters
near RLQs is $2-5$ times greater than the field at similar
redshifts. This suggests that at $z \gtrsim 0.8$, quasars are found in
regions of above-average star formation activity. There is also a
variation in the strength and number of ELG candidates found about
quasars, \eg MRC~B2037--234 has a number of line emitters consistent
with no overdensity in contrast to the clear clustering of strong ELG
candidates near MRC~B2156--245.

To estimate the SFR from [O\,{\sc ii}] luminosity we use the empirical
conversion of Gallagher \etal (1989)\nocite{gallagher89}:

\[
  \mathrm{SFR} \approx 1 \: \msun \: \mathrm{yr}^{-1} \left(
  \frac{L(\mathrm{[OII]})}{10^{41} \: \mathrm{erg} \:
  \mathrm{s}^{-1}} \right)
\]

\noindent
This conversion is uncertain by a factor of a few due to the scatter
in the samples used to calibrate the relationship. However, most
estimators of SFR are reasonably insecure because the estimates are
highly model dependent and include assumptions about the IMF,
metallicity and extinction (see the review by Kennicutt
1998\nocite{kennicutt98}).

\begin{figure}

  \epsscale{1}\plotone{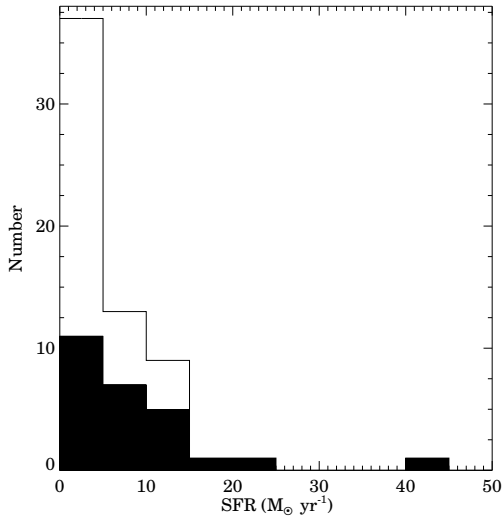}

  \figcaption[Barr.fig18.eps]{Histogram of star formation rates for
  ELG candidates in this work inferred from [O\,{\sc ii}] line
  luminosities using the relationship of Gallagher \etal (1989). The
  unfilled parts of the histogram represent those objects without a
  continuum magnitude whose SFR is a lower limit.\label{fig:sfr}}

\end{figure}

The resulting distribution of star formation rates is shown in
Figure~\ref{fig:sfr}. The SFRs span a range of $\sim 1 - 50$ \msun
yr$^{-1}$ with a median value of $4$ \msun yr$^{-1}$, that is also
consistent with those of field galaxies at similar redshifts
\citep{cowie97,hogg98,hicks02}.

Caution must be exercised in interpreting the above results. Vagaries
in the basic properties of ELGs at $z \sim 1$ such as gas ionization
and extinction make it difficult to compare samples selected using
different methods. The empirical conversion from $L$([O\,{\sc ii}])
to SFR is uncertain, and there is variation in the number of ELG
candidates found per field, as well as a small number found
overall. For these reasons, any quantitative analysis of the
luminosity function and star formation activity is likely to be
equivocal.

\subsection{Line emitters about AGN at other redshifts}

Our data provide a comparison with the detections of line emitters of
different types in the fields of AGN at other redshifts. Surveys of
this type vary in depth and sensitivity but yield surprisingly
consistent results. The [O\,{\sc ii}] narrow-band survey of Hutchings \etal
(1993)\nocite{hutchings93} of the fields of seven quasars and radio
galaxies at $z \sim 1.1$ found a density of $0.006 - 0.04$ Mpc$^{-3}$
ELG candidates, though the limiting SFR is not made explicit.

Teplitz, Malkan \& McLean (1998)\nocite{teplitz98} in their infrared 
search for H$\alpha$ emitters near $2.3 < z <2.5$ quasars found $0.0135 \pm
^{0.0055}_{0.0035}$ Mpc$^{-3}$. The average inferred SFR of these
candidates was 50 \msun yr$^{-1}$ and their density is $\sim 3$ times
that of field surveys at a similar redshift. Hall \etal
(2001)\nocite{hall01} also detected an overdensity of candidate
H$\alpha$ emitters about radio-loud quasars $\sim 3$ times over the
field, this time at $z \sim 1.5$.

The studies of Pentericci \etal (2000), Kurk \etal (2001) and Venemans
\etal (2002)\nocite{pentericci00,kurk01,venemans02} have found
tens of Ly$\alpha$ emitters around radio galaxies at $z=2.16$ and
$z=4.10$. The space density of these emitters is $0.01$ Mpc$^{-3}$ in
both cases and Kurk \etal claim to be sensitive to star formation
rates $\gtrsim 1$ \msun yr$^{-1}$. However, it remains notoriously
difficult to convert Ly$\alpha$ luminosity to SFR at $z > 2$ because
of the effect of intervening neutral hydrogen clouds. 

These studies show that powerful radio sources are typically found in
fields with overabundances of line emitters. Overdensities typically
range from $2 - 15$ times those of field surveys and velocity
dispersions are $\sim 300 - 1000$ km s$^{-1}$.

Our TTF study finds a space density of ELG candidates of $\sim 0.01$
Mpc$^{-3}$ with SFR $\gtrsim 1$ \msun yr$^{-1}$. The quasars detailed
here inhabit quantitatively similar environments to those at similar
and higher redshifts. Taken together these results suggest that
powerful radio sources trace similarly-overdense 
regions of active star-formation over a great range in redshift.

\section{Conclusions}

\begin{itemize}

\item We demonstrate that it is possible to isolate star-forming
galaxies at $0.8 < z < 1.3$ using tunable-filter observations. We have
detected forty-seven new ELG candidates in the fields of six
quasars. The candidates are selected on the basis of luminosity
changes across narrow wavelength ranges centred on redshifted [O\,{\sc ii}].

\item Radio-loud quasars at $0.8 < z < 1.3$ are found in regions of
above-average star formation activity. The number density of ELG
candidates about the quasars in our sample is $\sim 100$ times that of
local [O\,{\sc ii}] emitters and $2-5$ times the number found in
spectroscopic field surveys at $0.8 \lesssim z \lesssim 1.5$ at
$L$([O\,{\sc ii}]) $< 10^{42}$ erg s$^{-1}$.

\item On average, the space density and velocity distributions of ELG
candidates peak about the quasars. However, there is variance from
field to field. The number of candidates found in individual quasar
fields varies from that expected from field surveys, to overdensities
of $\sim 10$ times. 

\item The equivalent widths and inferred star formation rates of ELG
candidates in the quasar fields typically range between
$5<W_{\lambda}<100$\AA \ and $1<\mathrm{SFR}<50$ \msun yr$^{-1}$. The
median SFR is $4$ \msun yr$^{-1}$. These values are consistent with
those seen in the field at the same redshifts.

\item The ELG candidate distributions, velocity dispersions and star
formation rates at $z \sim 1$ detailed in this paper are consistent
with studies of line emitters in the fields of AGN at $1 < z <
4$. Radio sources inhabit actively star-forming regions over a wide
range in redshift.

\end{itemize}

\section*{Acknowledgments}

JCB acknowledges the support of a Royal Society University Research
Fellowship, and also support from NASA through Hubble Fellowship grant
\#HF-01103.01-98A from the Space Telescope Science Institute, which is
operated by the Association of Universities for Research in Astronomy,
Inc., under NASA contract NAS5-26555. RWH acknowledges funding from
the Australian Research Council. This research has made use of the
NASA/IPAC Extragalactic Database (NED) which is operated by the Jet
Propulsion Laboratory, California Institute of Technology, under
contract with the National Aeronautics and Space Administration.


\begin{thebibliography}{}
\bibitem[Axon \etal 2000]{axon00}%
Axon, D.~J., Capetti, A., Fanti, R., Morganti, R., Robinson, A., \& Spencer, R. 2000, AJ, 120, 2284
\bibitem[Balogh \etal 1997]{balogh97}%
Balogh, M.~L., Morris, S.~L., Yee, H.~K.~C., Carlberg, R.G., \& Ellingson, E. 1997, ApJ, 488, L75
\bibitem[{Balogh \etal }1998]{balogh98}%
Balogh, M.~L., Schade, D., Morris, S.~L., Yee, H.~K.~C., Carlberg, R.G., \& Ellingson E. 1998, ApJ, 504, L75
\bibitem[{Baker \etal }1999]{baker99}%
Baker, J.~C., Hunstead, R.~W., Kapahi, V.~K., \& Subrahmanya, C.~R. 1999, ApJS, 122, 29
\bibitem[{Baker \etal }2001]{baker01}%
Baker, J.~C., Hunstead, R.~W., Bremer, M.~N., Bland-Hawthorn, J., Athreya, R.~M., \& Barr, J.~M. 2001, AJ, 121, 1821.
\bibitem[{Barr }2003]{barr03t}%
Barr, J.~M. 2003, PhD Thesis, University of Bristol
\bibitem[{Barr \etal }2003]{barr03}%
Barr, J.~M., Bremer, M.~N., Baker, J.~C., \& Lehnert, M.~D. 2003, MNRAS, 346, 229
\bibitem[{Bertin \& Arnouts }1996]{bertin96}%
Bertin, E., \& Arnouts, S. 1996, AASS, 117, 393 
\bibitem[{Bland-Hawthorn \& Jones }1998{\em a}]{bland-hawthorn98a}%
Bland-Hawthorn, J., \& Jones, D.~H. 1998a, PASA, 15, 44
\bibitem[{Bland-Hawthorn \& Jones }1998{\em b}]{bland-hawthorn98b}%
Bland-Hawthorn, J., \& Jones, D.~H. 1998b, Proc. SPIE, Optical Astronomical Instrumentation, Sandro D'Odorico; Ed., 3355, 855
\bibitem[{Boyle \& Terlevich }1998]{bt98}%
Boyle, B.~J., Terlevich, R.~J., 1998, MNRAS, 293, L49 
\bibitem[{Bremer \etal }1992]{bremer92}%
Bremer, M.~N., Crawford, C.~S., Fabian, A.~C., \& Johnstone, R.~M. 1992, MNRAS, 254, 614 
\bibitem[{Bremer \etal }2002]{bremer02}%
Bremer, M.~N., Baker, J.~C., \& Lehnert, M.~D. 2002, MNRAS, 337, 470
\bibitem[{Butcher \& Oemler }1984]{butcher84}%
Butcher, H., \& Oemler, A. 1984, ApJ, 285, 426
\bibitem[{Cardiel \etal }2003]{cardiel03}%
Cardiel, N., Elbaz, D., Schiavon, R.~P., Willmer, C.~N.~A., Koo, D.~C., Phillips, A.~C., \& Gallego, J. 2003, ApJ, 584, 76 
\bibitem[{Cowie \etal }1997]{cowie97}%
Cowie, L.~L., Hu, E.~M., Songaila, A., \& Egami E. 1997, ApJ, 481, L9
\bibitem[{Cowie, Songaila \& Barger }1999]{cowie99}%
Cowie, L.~L., Songaila, A., \& Barger, A.~J. 1999, AJ, 118, 603
\bibitem[{Dressler \& Gunn }1992]{dressler92}%
Dressler, A., \& Gunn, J.~E. 1992, ApJS, 42, 565
\bibitem[Eales 1985]{eales85}
Eales, S.~A. 1985, MNRAS, 217, 149
\bibitem[{Ellingson, Yee \& Green }1991]{ellingson91b}%
Ellingson, E., Yee, H.~K.~C., \& Green, R.~F. 1991, ApJ, 371, 49
\bibitem[{Ellis \etal }1996]{ellis96}%
Ellis, R.~S., Colless, M., Broadhurst, T., Heyl, J., \& Glazebrook K. 1996, MNRAS, 280, 235
\bibitem[{Finn, Impey, \& Hooper }2001]{finn01}%
Finn, R.~A., Impey, C.~D., \& Hooper, E.~J.\ 2001, \apj, 557, 578 
\bibitem[{Gallagher, Hunter \& Bushouse }1989]{gallagher89}%
Gallagher, J.~S., Hunter, D.~A., \& Bushouse, H. 1989, AJ, 97, 700
\bibitem[Gallego \etal 1995]{gallego95}%
Gallego, J., Zamorano, J., Arag{\'o}n-Salamanca, A., \& Rego, M. 1995, ApJ, 455, L1
\bibitem[Gallego \etal 2002]{gallego02}%
Gallego, J., Garc{\' \i}a-Dab{\' o}, C.~E., Zamorano, J., Arag{\'o}n-Salamanca, A., \& Rego, M. 2002, ApJ, 570, L1
\bibitem[{Grazian \etal }2000]{grazian00}%
Grazian, A., Cristiani, S., D'Odorico, V., Omizzolo, A., \& Pizzella, A. 2000, AJ, 119,2540
\bibitem[{Hall \& Green }1998]{hall98}%
Hall, P.~B.~\& Green, R.~F.\ 1998, \apj, 507, 558 
\bibitem[{Hall \etal }2001]{hall01}%
Hall, P.~B., \etal 2001, AJ, 121, 1840 
\bibitem[{Hammer \etal }1997]{hammer97}%
Hammer, F., \etal 1997, ApJ, 481, 49
\bibitem[Hicks \etal 2002]{hicks02}%
Hicks, E.~K.~S., Malkan, M.~A., Teplitz, H.~I., McCarthy, P.~J., \& Yan, L. 2002, ApJ, 581, 205
\bibitem[Hogg \etal 1998]{hogg98}%
Hogg, D.~W., Cohen, J.~G., Blandford, R., \& Pahre, M.~A. 1998, ApJ, 504, 622
\bibitem[Hutchings \etal 1993]{hutchings93}%
Hutchings, J.~B., Crampton, D., \& Persram, D. 1993, AJ 106, 4
\bibitem[Hutchings \etal 2001]{hutchings01}%
Hutchings, J.~B., Morris, S.~L., \& Crampton, D. 2001, AJ, 121, 80
\bibitem[{Jones \& Bland-Hawthorn }2001]{jones01}%
Jones, D.~H., \& Bland-Hawthorn, J. 2001, ApJ, 550, 593 
\bibitem[{Jones \etal }2002]{jones02}%
Jones, D.~H., Shopbell, P.~L., \& Bland-Hawthorn, J. 2002, MNRAS, 329, 759
\bibitem[{Kapahi \etal }1998]{kapahi98}%
Kapahi, V.~K., Athreya, R.~M., Subrahmanya, C.~R., Baker, J.~C., Hunstead, R.~W., McCarthy, P.~J., \& van Breugel, W. 1998, ApJS, 118, 327
\bibitem[{Kennicutt }1992]{kennicutt92}%
Kennicutt, R.~C. 1992, ApJ, 388, 310
\bibitem[{Kennicutt }1998]{kennicutt98}%
Kennicutt, R.~C. 1998, ARA\&A, 36, 189
\bibitem[{Khochfar \& Burkert }2001]{khochfar01}%
Khochfar, S., \& Burkert, A. 2001, ApJ, 561, 517 
\bibitem[{Kurk \etal }2001]{kurk01}%
Kurk, J.~D., Pentericci, L., R{\" o}ttgering, H.~J.~A., \& Miley, G.~K. 2001, Astrophysics and Space Science Supplement, 277, 543
\bibitem[{Lacey \& Cole }1993]{lacey93}%
Lacey, C., \& Cole, S. 1993, MNRAS, 262, 627 
\bibitem[{Laing \etal }1983]{laing83}%
Laing, R.~A., Riley, J.~M., \& Longair, M.~S. 1983, MNRAS, 204, 151
\bibitem[{Lilly \etal }1996]{lilly96}%
Lilly, S.~J., Le Fevre, O., Hammer, F., \& Crampton, D. 1996, ApJ, 460, L1
\bibitem[{Madau \etal }1996]{madau96}%
Madau, P., Ferguson, H.~C., Dickinson, M.~E., Giavalisco, M., Steidel, C.~C., \& Fruchter, A. 1996, MNRAS, 283, 1388
\bibitem[{Madau, Pozzetti \& Dickinson }1998]{madau98}%
Madau, P., Pozzetti, L., \& Dickinson, M. 1998, ApJ, 1998, 498
\bibitem[{McCarthy }1993]{mccarthy93}%
McCarthy, P.~J. 1993, ARA\&A, 31, 639
\bibitem[{McLure \& Dunlop }2001]{mclure01}%
McLure, R.~J., \& Dunlop, J.~S. 2001, MNRAS, 321, 515 
\bibitem[{Miller \& Owen }2003]{miller03}%
Miller, N.~A., \& Owen, F.~N. 2003, AJ, 125, 2427 
\bibitem[{Murali \etal }2002]{murali02}%
Murali, C., Katz, N., Hernquist, L., Weinberg, D.~H., \& Dav{\' e}, R. 2002, ApJ, 571, 1
\bibitem[Murdoch \etal 1984]{murdoch84}%
Murdoch, H.~S., Hunstead, R.~W., \& White, G.~L. 1984, PASA, 5, 341
\bibitem[{Nakata \etal }2001]{nakata01} 
Nakata, F.~et al.\ 2001, \pasj, 53, 1139 
\bibitem[{Pentericci \etal }2000]{pentericci00}%
Pentericci, L., \etal 2000, A\&A, 361, L25
\bibitem[{Rhoads \etal }2000]{rhoads00}%
Rhoads, J.~E., Malhotra, S., Dey, A., Stern, D., Spinrad, H., \& Jannuzi, B.~T. 2000, ApJ,545, L85
\bibitem[Rush \etal 1997]{rush97}%
Rush, B., McCarthy, P.~J., Athreya, R.~M., \& Persson, S.~E. 1997, ApJ, 484, 163
\bibitem[Snellen \etal 1996]{snellen96}%
Snellen, I.~A.~G., Bremer, M.~N., Schilizzi, R.~T., Miley, G.~K., \& van Ojik, R. 1996, MNRAS, 279, 1294
\bibitem[Teplitz \etal 1998]{teplitz98}%
Teplitz, H.~I., Malkan, M., \& McLean, I.~S. 1998, ApJ, 506, 519
\bibitem[{Venemans \etal }2002]{venemans02}%
Venemans, B.~P., \etal 2002, ApJ, 569, L11
\bibitem[{Wang et al. }2004]{wang04}%
Wang, J.~X., et al.\ 2004, \apjl, 608, L21 
\bibitem[{Wilkes }1986]{wilkes86}%
Wilkes, B.~J.\ 1986, MNRAS, 218, 331 
\bibitem[{Wold \etal }2000]{wold00}%
Wold, M., Lacy, M., Lilje, P.~B., \& Serjeant, S. 2000, MNRAS 316, 267
\bibitem[{Wold \etal }2001]{wold01}%
Wold, M., Lacy, M., Lilje, P.~B., \& Serjeant, S.\ 2001, \mnras, 323, 231 
\bibitem[{Yee \& Green }1984]{yee84}%
Yee, H.~K.~C., \& Green, R.~F. 1984, ApJ, 280, 79
\bibitem[{Yee \& Ellingson }2003]{yee03}%
Yee, H.~K.~C., \& Ellingson, E. 2003, ApJ, 585, 215 

\end{thebibliography}
\end{document}